# Hybrid Improper Ferroelectricity and Moiré Superlattices-induced Exciton Quantization in Layered 2D Halide Perovskite

Sanika S. Padelkar[1,2,3,4,5], Sharidya Rahman[2,3 #], Mattia Belotti[1 #], Naufan Nurrosyid[2,3], Craig Forsyth[1], Alasdair Mckay[1], Tam Nguyen[1], Thi Vu Mung[1], Lan Nguyen[2,3], Naeimeh Mozaffari[2,3], Alexandr N. Simonov[1*], Aftab Alam[4,5*], Jacek J. Jasieniak[2,3*]

1. School of Chemistry, Monash University, Melbourne, Victoria, 3800, Australia
2. Department of Materials Science and Engineering, Monash University, Melbourne, Victoria, 3800, Australia
3. Centre of Excellence in Exciton Science, Monash University, Melbourne, Victoria, 3800, Australia.
4. Department of Physics, Indian Institute of Technology Bombay, Powai, Mumbai-400076
5. IITB-Monash Research Academy, Indian Institute of Technology Bombay, Powai, Mumbai-400076

# These authors contributed equally to this work.

* E-mail: jacek.jasieniak@monash.edu; alexandr.simonov@monash.edu; aftab@iitb.ac.in

## ABSTRACT

2D Ruddlesden-Popper perovskites are compelling platforms for quantum-confined optoelectronics. However, polar order in iodide composition remains rare under ambient conditions, and the mechanistic origin of anomalous photoluminescence in this class of perovskite is still speculative. Here, we demonstrate that solution-grown $(PA)_2FAPb_2I_7$ single crystals develop an inadvertent moiré superlattice through pseudo-merohedral twinning, driven by hybrid improper ferroelectricity in which trilinear mode coupling between two primary zone-boundary modes ($X_2^+$ and $X_3^-$) and a secondary $\Gamma_4^-$ polar displacement simultaneously breaks inversion symmetry and imposes a ca. 5.17° rotational misalignment between adjacent layers. This symmetry breaking activates one of the highest piezoelectric coefficients $d_{33}$ (ca. 20 pm/V) reported among 2D perovskites. This misalignment generates a moiré superlattice that undergoes a thermally driven commensurate-incommensurate transition, switching between a periodic confinement potential that quantizes excitons into an equidistant photoluminescence ladder at 123 K and a disordered incommensurate phase with broadened emission at 298 K. These emissions are attributed to moiré-confined excitons, resolving a longstanding debate on anomalous secondary photoluminescence in layered 2D perovskites and opening pathways to twistronics, photoferroelectrics and piezo-optoelectronic devices.

# MAIN

Low-dimensional hybrid organic inorganic perovskites (HOIPs) combine the energetic landscape of inorganic metal-halide octahedra with the lattice vibrations (phonons) of intercalated organic molecular chains and provide a unique platform with potential for overlapping emissive and vibrational properties. This complex interplay in 2D perovskites propels the formation of quasiparticles, like polarons, when photoexcited charge carriers (excitons) are strongly coupled with lattice fluctuations (phonons). Under optical excitation, such exciton-phonon coupling is significant due to dielectric confinement effects, high exciton binding energies, and highly anharmonic fluctuations in 2D perovskites[1]. This environment enables the study of rich and diverse photophysical mechanisms governing electron-phonon, exciton-phonon, exciton-polaron, and phonon-polariton interactions, as well as quasiparticle emissions such as excitons, biexcitons, trions, and phonon-assisted emissions[2,3]. The dressing of polaronic character to the exciton can impact the intrinsic properties of such 2D perovskite systems, viz. optical band-to-band transition, charge carrier mobility, non-radiative and radiative carrier recombination, dielectric response, and photoluminescence dynamics[4,5].

In this context, 2D Ruddlesden-Popper (RP) perovskite has documented emission spectra exhibiting several secondary peaks and often a Stoke-shifted low-energy broad photoluminescence (PL) emission, indicative of the overlap of vibrational (phonon) and emissive properties[6,7,8]. Recently, quasi-2D RP perovskite $BA_2MA_{n-1}Pb_nI_{3n+1}$ (for $n > 1$) exfoliated single crystals (where, $BA^+$ is Butyl Ammonium and $MA^+$ is Methylammonium) have attracted significant interest due to their surface edge states (secondary PL emission), which anomalously dissociate excitons into long-lived free carriers[9], attributed to either crystal lattice distortion[10] or spontaneous 3D perovskite formation[11,12] at crystal edges. Hong et al.[13], and Qin et al.[14] correlate edge states with non-centrosymmetry, where internal electric fields from polarized A-site cations drive ferroelastic stress-induced edge deformation, enabling strong piezoelectric coupling between electric fields and phonon-driven mechanical deformations. Qin et al.[14] further confirmed coexisting ferroelectric interior and paraelectric edge phases in $BA_2CsPb_2Br_7$, linking structural phase dynamics to secondary emission.

A complementary route to engineering the excitonic landscape of 2D materials is through moiré superlattices, formed when two periodic lattices are overlaid with a small relative twist angle. Lattice mismatch, twist angle, or strain along the layers of 2D van der Waals (vdW) - layered materials can create moiré superlattices, which induces flat bands and strong electron correlations due to the periodic potential imposed by the superlattice. The periodic potential modulates the electronic structure and exciton landscape in these heterostructures. This modulation leads to the formation of new quantum states, including multiple exciton energy levels and minibands, which manifest as additional PL peaks beyond those found in the constituent monolayers. Spectroscopic measurements in $WSe_2/WS_2$ and $MoSe_2/WS_2$ heterostructures reveal multiple PL peaks, whose energies and intensities depend on the twist angle, doping, and other tunable parameters, directly attributable to hybridization and miniband formation due to the superlattice[15]. In some systems, such as copper nanocluster assemblies[16], new quantum emissive states appear due to interlayer interactions enabled by the moiré pattern, resulting in "tunable twin emissions", which is a direct signature of secondary emission induced by new quantum states inherent to the superlattice[17].

Beyond the traditional vdW-layered materials, halide perovskite superlattices offer an exciting avenue for building twisted layers. When two ultrathin halide perovskite nanosheets (such as $MAPbI_3$ or other 2D lead halide perovskites) are overlaid with a small relative twist angle, a periodic superlattice emerges[17,18]. The ionic interlayer coupling in 2D perovskites is inherently stronger than the vdW forces governing graphene or TMD stacking, suggesting more robust moiré-induced quantum states. However, precise twist-angle control in layered perovskites remains elusive, and naturally occurring moiré superlattices in solution-grown perovskite crystals have not previously been identified as the source of anomalous emissive behaviour[17].

Inspired by previous findings and knowledge gaps in the origin and properties of the PL anomalies in quasi-2D RP perovskites[19], we investigated the implications of structure dynamics, electron-phonon coupling, and piezoelectricity on anomalous photoluminescence dynamics in these materials. Such a study requires a quasi-2D RP perovskite with room-temperature (ca. 298 K) piezoelectricity. However, most of them have Curie temperatures below ambient and the few known exceptions exhibit bandgaps above 2.5 eV, precluding their use in visible-range ferroelectric and electro-optic applications[20].

To address the above challenges, we introduce $(PA)_2FAPb_2I_7$ single crystal ($PA^+$ is pentylammonium and $FA^+$ is formamidinium), as one of the first iodide-based quasi-2D RP perovskites to exhibit piezoelectric behaviour under ambient conditions. Through combined structural, spectroscopic, and theoretical investigation, we demonstrate that a hybrid improper ferroelectric mechanism and spontaneous pseudo-merohedral twinning during the crystal growth, imposes a rotational misalignment between adjacent layers that generates a moiré superlattice. This superlattice undergoes a thermally driven commensurate–incommensurate transition, and we establish that the anomalous secondary photoluminescence characteristic of this material originates from excitons quantum-confined within the resulting periodic moiré potential. Thus, resolving a longstanding mechanistic ambiguity and directly linking structural symmetry breaking to quantized optical emission in solution-grown layered perovskites.

# RESULTS

Micrometre to millimetre-sized single crystal flat flakes of $(PA)_2FAPb_2I_7$ were grown using a solution temperature-lowering (STL) method[21]. A controlled thermal gradient and an optimized reaction mixture composition were required to facilitate gradual nucleation and single crystal growth (Figure 1a). The crystals were further characterised through a comprehensive set of characterisation techniques (see Methods) including structural (X-ray diffraction), compositional (microscopy), electromechanical (piezoelectric response), optical (spectroscopy), as well as theoretical and computational analysis.

## Symmetry Breaking, Lattice Reconstruction and Phonon Dynamics

Single-crystal X-ray diffraction (SCXRD) analysis revealed that the as-synthesized $(PA)_2FAPb_2I_7$ single crystals belong to the 2D RP perovskite family (Supplementary Table 1). The crystal structure of $(PA)_2FAPb_2I_7$ in orthorhombic phase represents a sequence of stacked 3D $FAPbI_3$ perovskite-like layers ($n$ = 2) intercalated by the $PA^+$ layers (Supplementary Figure 2). Compositional homogeneity of $(PA)_2FAPb_2I_7$ is substantiated by elemental mapping of exfoliated $(PA)_2FAPb_2I_7$ single crystals using scanning electron microscopy coupled X-ray energy dispersive spectroscopy (SEM-EDS) (Figure 1b). Inductively coupled plasma mass spectrometric (ICP-MS) analysis (Supplementary Note 1; Supplementary Figure-1a-b) further unveils that synthesized $(PA)_2FAPb_2I_7$ contains ca. 26.4 at. % Pb and ca. 57.1 at. % I, which is comparable to the theoretical values of 27.2 at. % and 58.4 at. %, respectively. In turn, $^1H$ and $^{13}C$ nuclear magnetic resonance (NMR) spectra of the solution of $(PA)_2FAPb_2I_7$ in deuterated dimethylsulphoxide (DMSO-$d_6$) confirm the expected $PA^+$ : $FA^+$ ratio of 2:1 (Figure 1c,d; Supplementary Note 2).

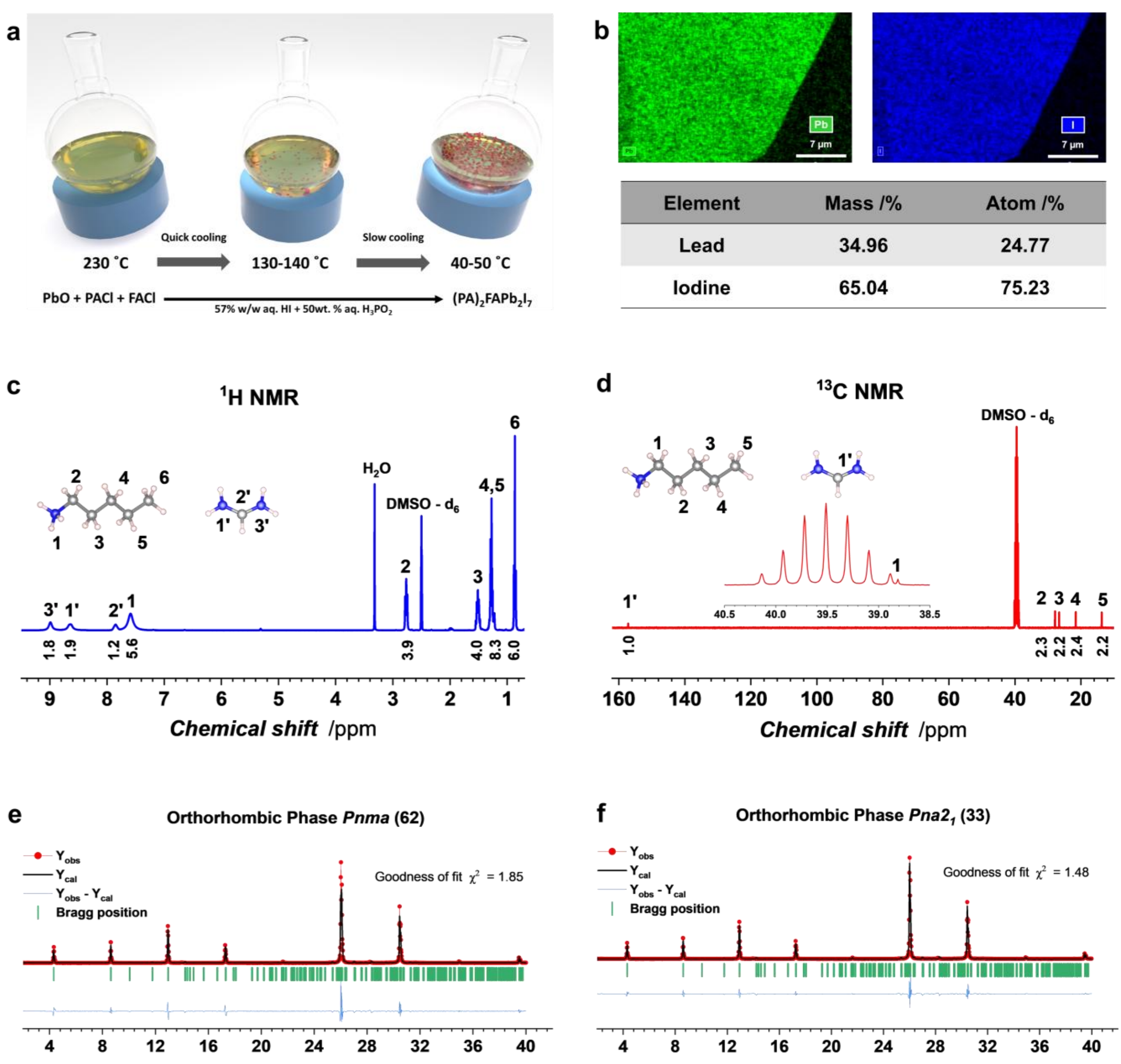

| Element | Mass /% | Atom /% |
|---|---|---|
| Lead | 34.96 | 24.77 |
| Iodine | 65.04 | 75.23 |

**Figure 1. Synthesis of the $(PA)_2FAPb_2I_7$ single crystals.**
(a) Schematic presentation of the experimental setup and conditions used for the synthesis. (b) SEM-EDS elemental mapping of the exfoliated $(PA)_2FAPb_2I_7$ single crystal along with the atomic concentrations of Pb and I. (c) $^1$H and (d) $^{13}$C NMR spectra of the $(PA)_2FAPb_2I_7$ solution in DMSO-$d_6$ highlighting the signals associated with $PA^+$ and $FA^+$; labels above and below signals show corresponding assignments and integrals. (e-f) Experimental PXRD pattern (*red*) compared to simulated data based on the centrosymmetric space group non-centrosymmetric (e) Pnma and (f) $Pna2_1$ (*black*), the experiment-simulation difference (*blue*), and Bragg positions (*green*).

Phase purity and high crystallinity were verified by Room temperature (RT) Powder X-ray diffraction (PXRD) measurements, with characteristic low-angle reflections at 4.35° and 8.70° confirming the $n$ = 2 Pb-I octahedral layer periodicity (Figure 1e-f). A central observation is that both *Pnma* (No. 62) and *Pna*2₁ (No. 33) gave statistically comparable fits to the experimental PXRD pattern (Figure 1e-f). This near-equivalence reflects the group-subgroup relationship between the two space groups, wherein *Pna2₁* is a maximal non-centrosymmetric subgroup of *Pnma*. Since the primitive lattice translations and glide-plane operations are retained across this symmetry reduction (Supplementary Note 3), the systematic absences and peak positions are nearly indistinguishable between the two models. Le Bail fitting[22] marginally favoured $Pna2_1$, though diffraction intensities alone cannot resolve the ambiguity. A refined and coherent structural description is discussed subsequently.

Differential scanning calorimetry (DSC) (Supplementary Figure 3b) identified a well-defined thermal anomaly at $T_c$ *ca.* 349 K, signalling a structural phase transition well above RT. This is consistent with the PXRD refinements, which favor the non-centrosymmetric $Pna2_1$ model at RT, suggesting that inversion symmetry breaking, if present, occurs under ambient conditions. Variable temperature high resolution PXRD examined between 123 K and 373 K confirmed a smooth, monotonic lattice expansion below $T_c$, with no evidence of reconstructive structural transition. Above $T_c$, a second paraelectric phase emerged, identifiable by the appearance of an additional reflection peak shifted by ca. 0.5° towards lower 2θ (Supplementary Figure 2a,c).

SCXRD refinements at RT showed that the diffraction data could be satisfactorily modelled in either *Pnma* or $Pna2_1$, which are related by a group-subgroup relationship (Figure 2a-b; Supplementary Table 1). The two space groups differ principally in the mirror plane $m_z$(i.e., $\sigma_{xy}$) at y = ¼ or ¾ present in *Pnma*, running through the $FA^+$ cation and one iodine atom, (dashed square marked in Figure 2a(ii)-b(ii); Supplementary Note 3). Once pseudo-merohedral twinning and $PA^+$ orientational disorder are taken into account, *Pnma* emerges as the more probable average structure, while the possibility of local $Pna2_1$ regions is not excluded (See Supplementary Note 4 for details). The necessity of invoking a twin law indicates that the crystal contains multiple symmetry-equivalent structural variants related by rotational operations. Such twinning leads to averaging of weak inversion-breaking displacements within a coherently diffracting domain. This naturally biases the refinement towards an apparently centrosymmetric *Pnma* model. This highlights the inherent limitation of SCXRD in resolving subtle or spatially varying non-centrosymmetry when rotational variants are present (Supplementary Note 4).

The structural relationship between *Pnma* and $Pna2_1$ was analysed using group-theoretical symmetry-mode decomposition. The *Pnma* to $Pna2_1$ transition can proceed along two symmetry-allowed pathways: a zone-centre polar mode $\Gamma_4^-$ associated with proper ferroelectricity, or two zone-boundary non-polar rotational $X_2^+$ and tilt $X_3^-$ modes, that act as the primary order parameters and induce secondary polarisation through trilinear mode coupling (Supplementary Note 5; Supplementary Table 2 & 3). The latter pathway is the hallmark of hybrid improper ferroelectricity (HIF), well documented in layered oxide perovskites such as $(Ca,Sr)_3Ti_2O_7$ [23,24].

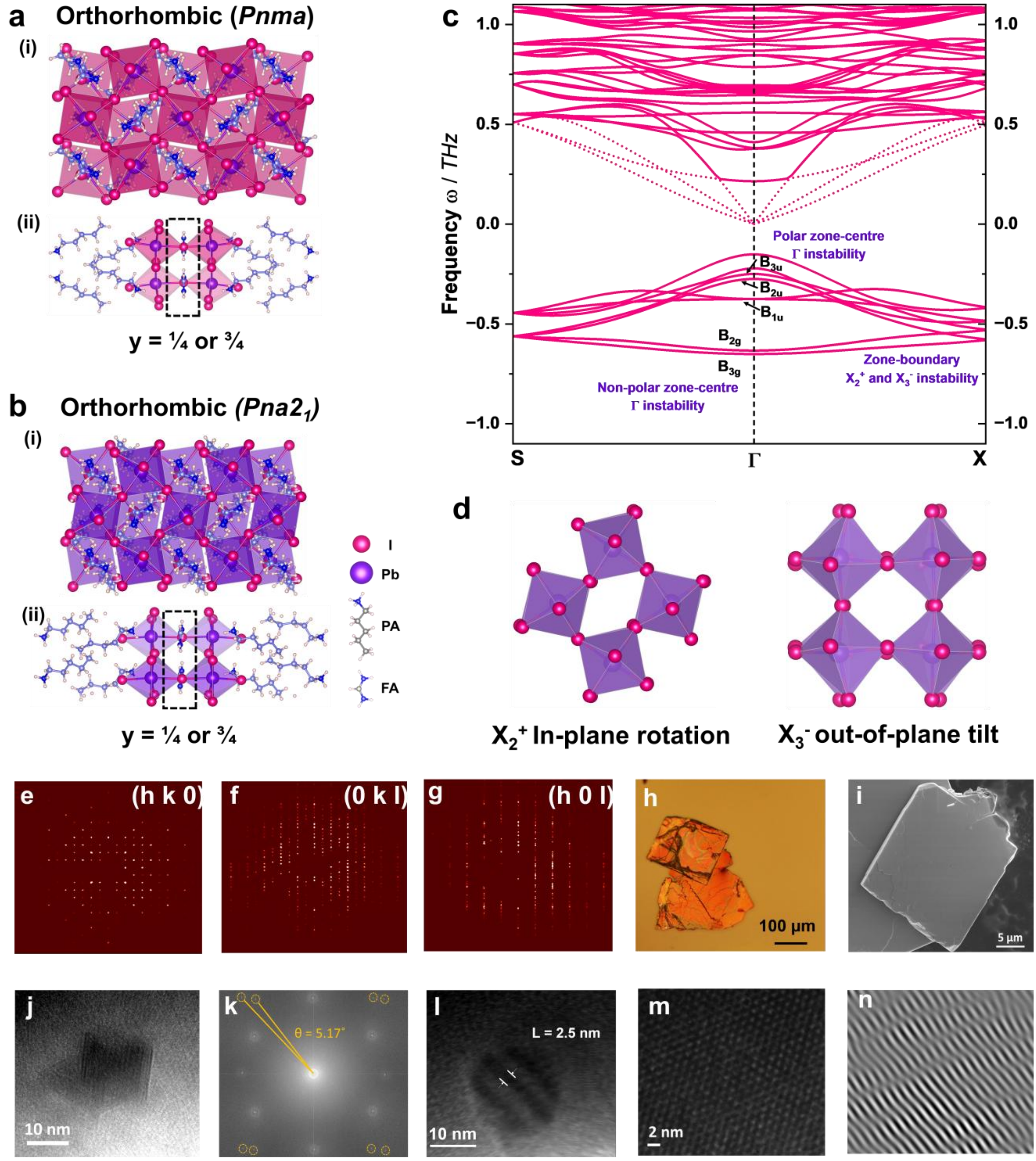

**Figure 2. Phase transitions and Phonon dynamics in $(PA)_2FAPb_2I_7$.**
Crystal structure of $(PA)_2FAPb_2I_7$ visualized in a **(a)** centrosymmetric structure (Orthorhombic Pnma) and **(b)** non-centrosymmetric structure (Orthorhombic $Pna2_1$) respectively. **(c)**, Simulated Phonon spectrum of $(PA)_2FAPb_2I_7$ in Pnma phase with phonon instabilities due to zone-center polar displacement mode ($\Gamma_4^-$) at $k = (0,0,0)$, and zone-boundary ($X_2^+$) rotational mode and ($X_3^-$) tilt mode at, $k = (1/2,0,0)$ which is non-polar by itself, but can lead to hybrid improper ferroelectricity. **(d)** Pnma to $Pna2_1$ phase transition of $(PA)_2FAPb_2I_7$ due to cooperative octahedral rotations ($X_2^+$) and tilting ($X_3^-$), corresponding to a zone-boundary $X_2$ mode rotational misalignment/distortion **(e-g)** Precession images of $(PA)_2FAPb_2I_7$ simulated from SCXRD along (hk0) with a grid of sharp discrete spots, indicating inter-layer alignment. However, along the (0kl) and (h0l) planes, streaking or elongation of diffraction spots signifies disorder or misalignment between layers. **(h)** Optical microscopy image revealing pseudo-merohedral twinning in single crystals. **(i)** SEM images of single crystal. **j**, TEM image of single crystal revealing ferroelastic twin domains **(k)** SAED pattern of the single crystal flakes, indicating misalignment in layers with a relative twist angle ($\theta$) = 5.17° **(l)** Bright-field TEM image of $(PA)_2FAPb_2I_7$ single crystal flakes, exhibiting moiré pattern with moiré periodicity L = 2.5 nm. **(m)** Zoom-in HRTEM image of the selected moiré pattern regime illustrating the interference lattices. **(n)** Inverse Fast Fourier Transform (IFFT) image obtained from selected diffraction spots exhibit noticeable wavy non-uniform lattice fringes.

The lattice dynamical properties of the high-symmetry *Pnma* phase were analysed using density functional theory (DFT) to identify the distortion mode pathway leading to the polar *$Pna2_1$* phase. Phonon dispersion calculations (Figure 2c) revealed the presence of imaginary frequencies at both the Brillouin-zone center (Γ) and zone-boundary points (S and X), particularly the X point. Irreducible representation analysis at the Γ point identified weak polar instabilities ($B_{1u}$, $B_{2u}$, $B_{3u}$) alongside a dominant non-polar gerade mode ($B_{2g}$, $B_{3g}$), indicating that Γ-point instabilities alone cannot account for the *Pnma* to *$Pna2_1$* symmetry reduction; instead, the strongly unstable zone-boundary $X_2^+$ and $X_3^-$ modes at the X point constitutes the primary structural instability (Supplementary Note 5, Supplementary Table 2 & 3). The $X_2^+$ and $X_3^-$ modes distortion acts as the dominant, non-polar primary order parameter that indirectly breaks inversion symmetry through trilinear coupling to the secondary $\Gamma_4^-$ polar mode, a hierarchy that is the defining signature of HIF and explains the stabilisation of the polar *$Pna2_1$* phase.

In $(PA)_2FAPb_2I_7$, the $FA^+$ cations are dynamically disordered at RT, and even modest reorientations can couple to cooperative octahedral rotations through hydrogen bonding and steric interactions (Figure 2d). This coupling drives the condensation of the $X_2^+$ rotational and $X_3^-$ tilt modes, lifts the mirror symmetry, and ultimately stabilises the polar *$Pna2_1$* phase through trilinear coupling to the secondary $\Gamma_4^-$ polar mode, offering a chemically intuitive route to HIF in a layered 2D halide perovskite.

Laue precession images recorded along the principal reciprocal-space planes provided direct evidence for interlayer rotational disorder (Figure 2e-g). Sharp, discrete spots were observed along (*hk*0), indicating good in-plane order, whereas streak-like elongation of reflections along (*h*0*l*) and (0*kl*) revealed reduced coherence along the [010] stacking direction (Supplementary Note 6). This anisotropic diffuse scattering is the expected signature of local $X_2^+$-type rotational distortions that preserve in-plane periodicity while disrupting long-range stacking order. Optical microscopy revealed twin-like domain contrast within individual crystals, and SEM imaging confirmed that the crystals retain sharp facets, ruling out surface degradation as an artefact (Figure 2h,i).

At the nanoscale, high resolution transmission electron microscopy (HRTEM) imaging resolved coherent twin domain boundaries, complementing the optical observations (Figure 2j). Selected area electron diffraction (SAED) patterns displayed secondary diffraction spots offset from the primary lattice by a relative rotation of approximately 5.17°, indicative of weak misalignment between adjacent domains[25] (Figure 2k). From this twist angle, a moiré periodicity of ca. 2.55 nm was calculated using $L = a/(2\sin(\theta/2))$, in excellent agreement with the modulation of ca. 2.5 nm measured directly from HRTEM images[1526] (Figure 2l). This rotational mismatch gives rise to a moiré superlattice, indicating interference pattern arising from the overlay of two slightly misaligned periodic lattices (Figure 2l,m). Inverse FFT analysis further revealed spatially modulated, wavy lattice fringes consistent with localised strain and gradual twist-angle variation across twin domain boundaries (Figure 2n).

Overall, our structural phase analysis identifies $(PA)_2FAPb_2I_7$ as a phase-pure hybrid improper ferroelectric 2D RP perovskite, driving local $Pna2_1$ symmetry breaking, corroborated by nanoscale moiré superlattices arising from twin-domain rotational misalignment (Supplementary Note 6). Having defined this symmetry-driven structural dynamics, we next quantify the macroscopic piezoelectric and ferroelectric responses of the material.

## Electromechanical Response: Emergent Hybrid Improper Ferroelectricity and Piezoelectric Coupling

To investigate the local polarisation behavior in $(PA)_2FAPb_2I_7$, piezo-force microscopy (PFM) has been utilized[26,27]. For these measurements, $(PA)_2FAPb_2I_7$ perovskite single crystals were mechanically exfoliated to bulk and atomically thin layers and transferred to conductive Si substrate using the vdW transfer method. Figure 3a,d show the atomic force microscopy (AFM) topographies of exfoliated perovskite flakes, indicating smooth and clean interfaces after the mechanical fabrication process. This ensures that vdW stacking can be leveraged to precisely control the layer thickness, from bulk to bilayer structures. Figure 3c-d shows the corresponding PFM amplitude images of bulk $(PA)_2FAPb_2I_7$ in vertical and horizontal direction, respectively. Distinct piezoelectricity can be detected in both out-of-plane and in-plane direction. Similar observations were also made from atomically thin layers (Figure 3e-f), although the piezo amplitude was reduced. PFM phase maps showing the polarisation direction of ferroelectric domains in Supplementary Figure 4 further validate the presence of piezoelectricity at room temperature. Repeatable results can be obtained across multiple samples (Supplementary Figure 5).

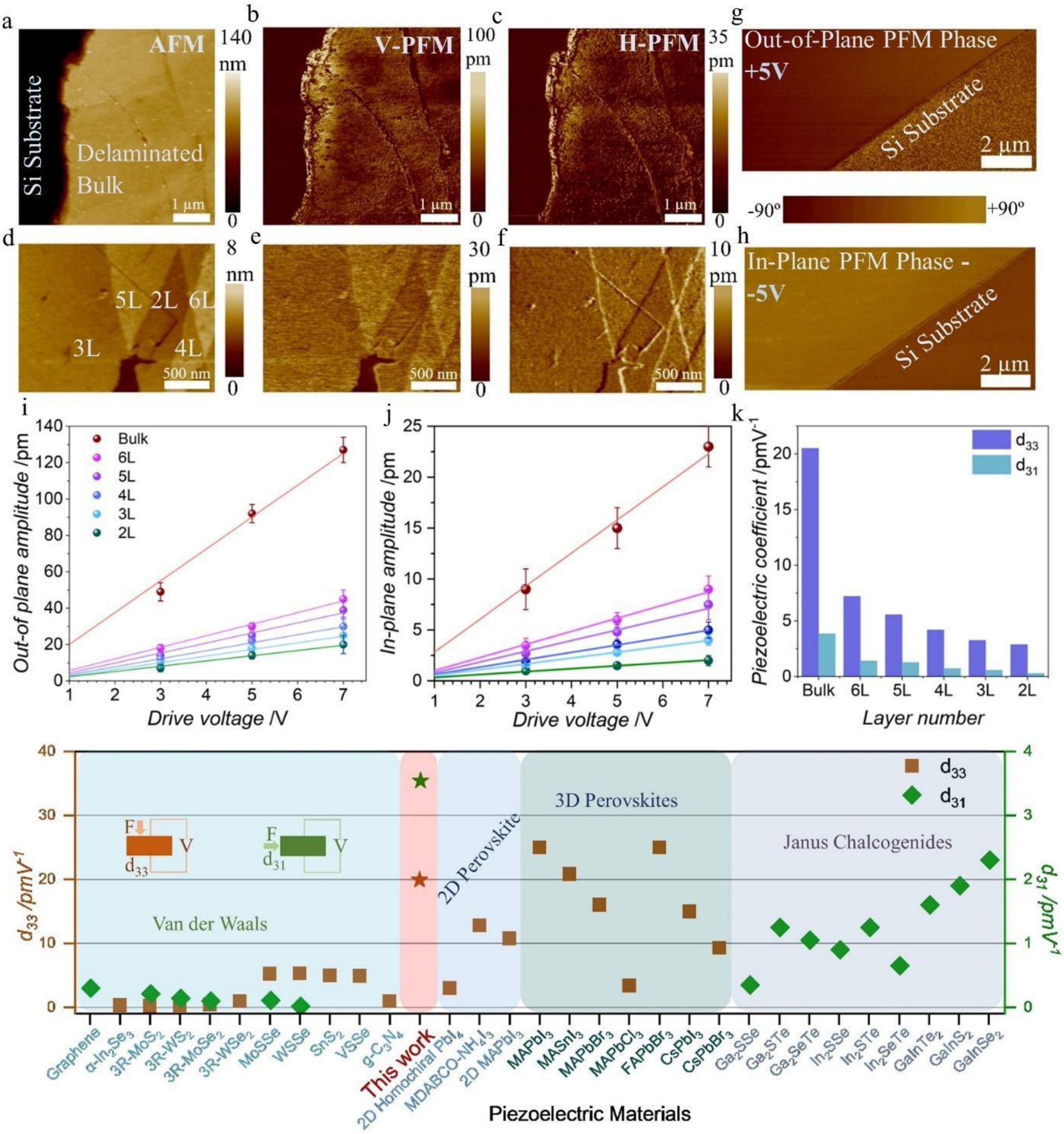

**Figure 3. Piezo force microscopy of $(PA)_2FAPb_2I_7$ and summary of piezoelectric coefficients. (a)** AFM topography of delaminated bulk $(PA)_2FAPb_2I_7$ perovskite on a Si substrate. Thickness is roughly about 100 nm. **(b-c)** Corresponding PFM amplitude along out-of-plane and in-plane direction respectively. Strong piezoelectricity can be witnessed in the vertical direction while weak but discernible piezo response can be ensured in the horizontal direction. **(d)** AFM topography of ultrathin $(PA)_2FAPb_2I_7$ perovskite cleaved mechanically in a similar Si substrate. Labelled layers

numbers indicate atomic thickness which can be obtained upon exfoliation. **(e-f)** Corresponding PFM amplitude in vertical and horizontal directions respectively. As witnessed, piezo response keeps decreasing with layer numbers. **(g-h)** Out-of-plane PFM phase images of another bulk sample taken at +5 and -5V respectively, upholding that switching of the orientation of the polar domains in the material is induced by changing the sign of the voltage bias applied to the tip. This is a typical behaviour of ferroelectricity. **(i-j)** Quantitative depiction of piezoelectric magnitude as a function of drive voltage along out-of-plane and in-plane direction respectively. Linear fitting is used to calculate the effective piezoelectric coefficient ($d_{33}$, $d_{31}$) from the slope. **(k)** Comparative depiction of out-of-plane ($d_{33}$) and in-plane ($d_{31}$) piezoelectric coefficients as a function of layer number of 2D $(PA)_2FAPb_2I_7$. **(l)** A comparative summary of piezoelectric coefficients for different state-of-the-art systems available in the literature. Y-axis represents the absolute values of $d_{33}$ and $d_{31}$. Note that some of these results are experimentally measured while others are obtained from first principal calculations. Extended comparison can be found in Table 4 of Supporting Information.

A characteristic feature of all ferroelectrics is the reorientation of polarisation or ferroelectric domain direction upon switching of the external electric field[27,28]. Such behavior can be clearly realized from the out-of-plane phase maps shown in Figure 3g-h, where the orientation of ferroelectric domain completely reverses upon changing the sign of the applied external voltage. Further quantitative analysis is shown in Supplementary Figure 6, where the signal profile across selected portions of the flakes is compared. PFM topography, amplitude and phase at +5V of another exfoliated flake match each other as shown in Supplementary Figure 6a-c. When the polarisation bias was reverted, the phase profile showed an opposite trend in the vertical direction (Supplementary Figure 6d,f); this is true down to atomically thin layers (Supplementary Figure 7). However, this observation is absent in the lateral direction where no change in phase direction can be detected (Supplementary Figure 6e,g). Overall, these measurements confirm out-of-plane ferroelectric effect in 2D $(PA)_2FAPb_2I_7$. To eliminate unwanted effects such as electrostatic, mechanical or chemical artifacts[29,30], a combination of electrostatic blind spot method, substrate subtraction method and off-resonance imaging have been used (Refer to Supplementary Note 7 and Figure 8 & 9)[31,32,33,34].

Piezoelectric properties of 2D perovskite were further analyzed by estimating the piezoelectric coefficients vertically ($d_{33}$) and horizontally ($d_{31}$)[35]. PFM scanning of bulk and atomically thin layer flakes were carried out at progressively increasing drive voltage. Figure 3i-j shows the average piezo amplitude for bulk and thin layers, along vertical and horizontal direction respectively at different bias voltages. The $d_{33}$ and $d_{31}$ coefficients determined from the slope of the measured piezo-responses vs drive voltage are plotted as a function of layer thickness in Figure 3k. Large $d_{33}$ values of ~20 pm/V are obtained for delaminated bulk samples which gradually reduce as the layer number decreases. The $d_{31}$ is proportionately lower in magnitude but follows the same trend with sample thickness. The increase of piezoelectric response with thickness has been observed previously in conventional layered ferroelectric materials and was attributed to the decreased substrate constraint and reduction in contribution from substrate clamping effect[30,36,37]. A comparison of piezoelectric coefficients of some common 2D and 3D perovskite along with widely known vdW layered and Janus materials are included in Figure 3l and Supplementary Table 4. Enhanced $d_{33}$ coefficient obtained in delaminated bulk perovskite is notably elevated compared to traditional vdW materials, Janus chalcogenides and even some 3D perovskite.

Piezoelectric strength scales with polarisation and permittivity[38]. In halide perovskites, polarisation arises primarily from A-site cation orientational freedom and inversion-symmetry breaking via B-site off-centering. In 2D $(PA)_2FAPb_2I_7$, $FA^+$ spacer dynamic orientation toggles centrosymmetric vs non-centrosymmetric stacking (Figure 2a–b), thereby controlling the piezoelectric response[38,39,40,41]. Structural investigation reveals a trilinear coupling mechanism leading to hybrid improper ferroelectricity. In addition, the long $PA^+$ chains in $(PA)_2FAPb_2I_7$ enlarge the spacing between $PbI_6^{-4}$ octahedral sheets, increasing electromechanical compliance and contributes to the high piezoelectricity. However. as thickness decreases, polar-domain density and stability are reduced while depolarisation fields are amplified, collectively diminishing the PFM signal (Supplementary Note 9)[42,43,44].

## Moiré-Induced Exciton Quantization and Photoluminescence Dynamics

Layer-dependent PL analysis of as-synthesized single crystals (SC), exfoliated single crystals (ESC), and atomically thin layers (ATL) established that the low-energy broad emission commonly attributed to edge states originates predominantly from thinner layers at the crystal periphery, consistent with the known thickness sensitivity of edge-state formation in 2D RP perovskites[12] (Supplementary Note 10, Supplementary Figure 10 & 11). Given the superior spectral resolution of both primary exciton emission $X_1$ and the secondary emissions manifold in ESC samples at room temperature (Figure 4d), ESC samples were employed for all temperature- and polarisation-resolved PL measurements reported hereafter.

Temperature-dependent PL measurements on ESC samples, collected between 373 K and 83 K, revealed a rich spectral evolution (Figure 4a, Supplementary Figure 12 b,e). At RT a single sharp emission $X_1$ dominated the spectrum, behaving analogously to a zero-phonon line. On heating above the structural phase transition near 349 K, $X_1$ quenched and only a broad low-energy PL persisted, consistent with the entry into the paraelectric *Pnma* phase. On cooling below 293 K, secondary emission features appeared alongside $X_1$, growing in intensity until ca. 173-203 K, below which they diminished. At 243 K a distinct intermediate peak $X_2$ became resolvable, and further cooling to 93 K revealed the full manifold $X_3$, $X_4$, and $X_5$ (Figure 4a).

A particularly informative feature emerged at 83-93 K: the $X_1$ emission split into a closely spaced doublet $X_1$ and $X_1'$. Power-dependent PL at 93 K confirmed that this splitting became more pronounced at higher excitation densities (Figure 4b), pointing to contributions from biexcitons, trions, or exciton fine-structure splitting (FSS) arising from exchange interactions and dielectric confinement in the quasi-2D lattice, phenomena well documented in layered halide perovskites[45,46,47].

The strong thermal broadening of the secondary emission at elevated temperatures, and its resolution into an equidistant ladder of peaks at low temperature, implicated exciton-phonon coupling as a central factor governing the spectral lineshape. Bose-Einstein[48] fits to the temperature-dependent PL linewidth $\Gamma(T)$ for ESC samples yielded a LO phonon coupling constant $\Gamma_{LO}$ = 165 ± 28.0 meV (Figure 4c, Supplementary Note 11), with the acoustic contribution $\Gamma_{ac}$ converging to zero. This coupling strength is notably larger than values reported for related 2D perovskites[6,49,50] and increases further in atomically thin layers ($\Gamma_{LO}$ = 555.2 ± 98.4 meV) (Supplementary Figure 14a-b ), indicating a pronounced enhancement of exciton-phonon coupling with reduced dimensionality.

Conventional assignments of broad or multiple low-energy emissions in low-dimensional perovskites invoke defect/trap states, self-trapped excitons (STEs), or inter-system crossing (triplet-singlet) transitions. A critical assessment of each scenario against the observed data is provided in Supplementary Note 12; none adequately accounts with the experimental observations (Supplementary Figure 15). Power-dependent PL at room temperature showed linear intensity scaling for $X_1$ ($\kappa$ = 1.19 ± 0.09) and near-linear scaling for $X_5$ ($\kappa$ = 0.90 ± 0.05), confirming the intrinsic character of both emissions[51] (Figure 4d). At 93 K, $X_1$ remained super-linear ($\kappa$ = 1.39 ± 0.16) while $X_3$, $X_4$, and $X_5$ retained linear dependence, ruling out localised trap or STE origins for these features[51] (Figure 4e; Supplementary Note 12).

At 93 K, the PL spectrum is deconvoluted cleanly into five components: $X_1$, $X_2$, $X_3$, $X_4$, $X_5$ with equal energy spacings of approximately 145 meV (Figure 4f). Such a periodic ladder of emission lines is the characteristic optical signature of excitons confined within a spatially modulated periodic potential, rather than a random distribution of traps or STEs.

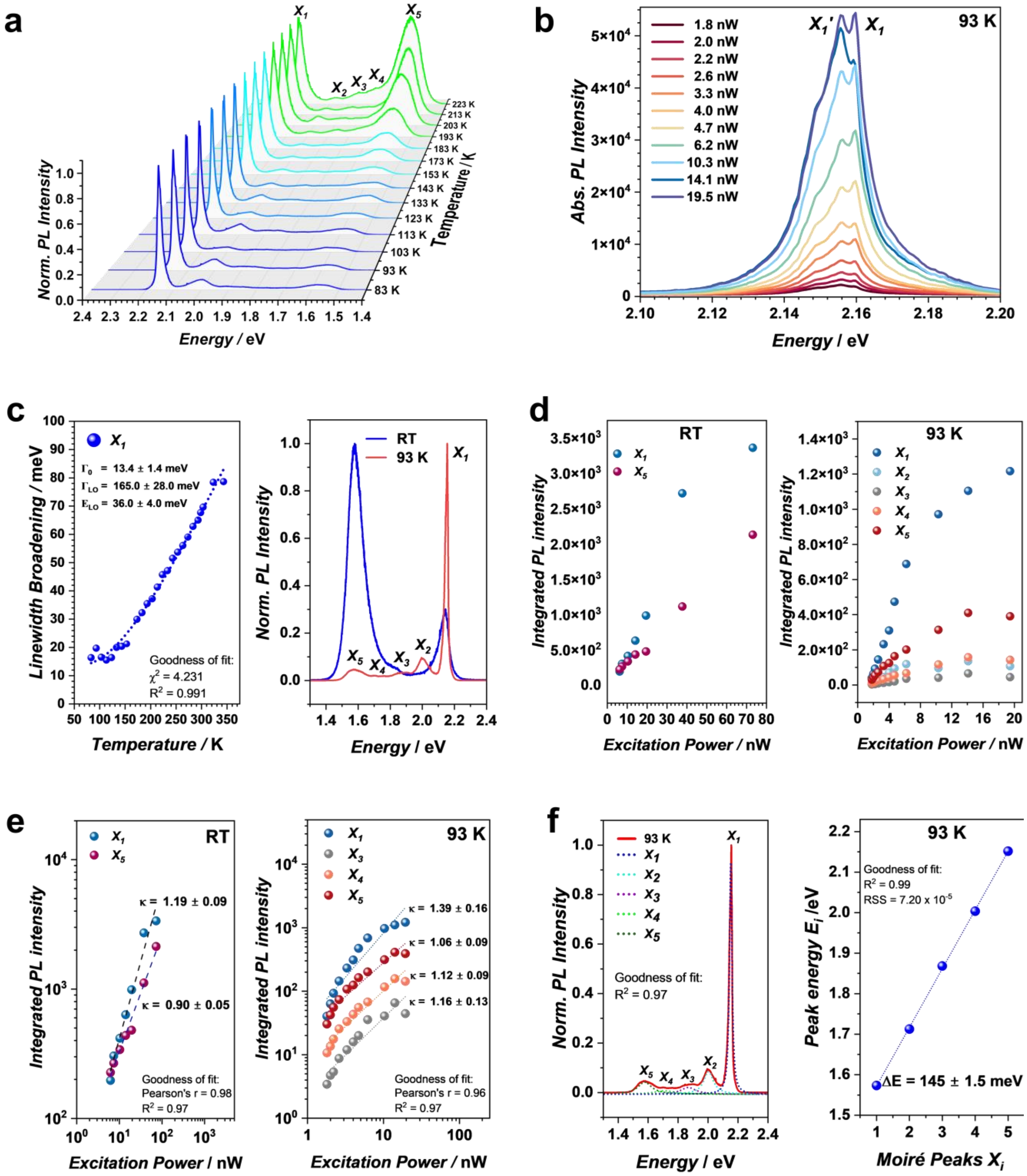

**Figure 4. Optical characterization and Exciton-phonon coupling in 2D $(PA)_2FAPb_2I_7$ perovskite flakes.**

**(a)** Temperature-dependent PL spectra of exfoliated single crystal at lower temperature, highlighting the emergence of equidistant PL emission peaks ($X_1$, $X_2$, $X_3$, $X_4$, and $X_5$) **(b)** Power-dependent PL spectra of exfoliated single crystal measured at 93 K indicating spectral splitting ($X_1$ and $X_1'$) **(c)** (left plot) Functional dependence of the PL linewidth $\Gamma$ (FWHM) of $X_1$ emission on temperature using equation (1) to estimate the strength of exciton scattering in exfoliated single crystal bulk. (right plot) PL spectra of exfoliated $(PA)_2FAPb_2I_7$ single crystal at RT and 93 K showing resolution of multiple equidistant peaks - $X_1$, $X_2$, $X_3$, $X_4$, and $X_5$. **(d)** Excitation power-dependent Integrated PL Intensity of $X_1$, $X_2$, $X_3$, $X_4$, and $X_5$ for exfoliated $(PA)_2FAPb_2I_7$ single crystal at RT (left) and 93 K (right). **(e)** Log-log plot of the PL intensity as a function of excitation power, I(PL) $\alpha$ $P^{\kappa}$ for $X_1$, $X_2$, $X_3$, $X_4$, and $X_5$ PL emissions for exfoliated $(PA)_2FAPb_2I_7$ single crystal at RT (left) and 93 K (right). **(f)** Deconvolution of PL spectra measured at 93 K into 5 PL peaks with equidistant spacing of around 145 meV.

The equidistant emission ladder finds a natural explanation within the moiré exciton paradigm, established by the microstructural analysis. In a moiré superlattice, excitons experience a spatially periodic potential $V_{moiré}$(r) that confines them within potential minima, generating quantised energy levels or minibands. The excitonic states are governed by the Hamiltonian H = $E_0$ + ($\hbar^2k^2/2M$) + $V_{moiré}$(r), where M is the exciton translational mass. Near the potential minima, harmonic confinement predicts equally spaced levels with ΔE ≈ $\hbar\omega$, and the observed spacing of ca. 145 meV is consistent with moiré potentials reported in other twisted vdW systems[24,52,53]. The series of equidistant PL peaks thus represents optical transitions from successive confined excitonic minibands within the moiré potential wells (Figure 4f).

Further two observations corroborate this assignment: (i) $X_5$ persisted as the sole observable emission above 350 K, well beyond the thermal stability of STE or defect emissions, which typically quench rapidly with temperature - demonstrating that the moiré trapping potential preserves excitonic integrity against thermal dissociation[54] (Figure 5a, Supplementary Note 12). (ii) polarisation-resolved PL measurements revealed a pronounced degree of linear polarisation (DOLP ≥ 0.6) for $X_5$, whereas $X_1$ displayed negligible anisotropy (Figure 5b-c). As the analyser angle was varied, the $X_5$ peak energy shifted measurably (Figure 5d), reflecting the energetic hierarchy of excitonic minibands with distinct polarisation selection rules imposed by the superlattice symmetry. Such a behaviour where angular energy shifts together with high and stable linear polarisation, is absent in STEs and defect emissions, which at most modulate PL intensity without shifting peak energy[55,56] (Supplementary Note 13).

Collectively, the temperature-dependent PL, power-dependent measurements, exciton-phonon coupling analysis, and polarisation-resolved spectroscopy converge on a consistent picture: the secondary emission ladder in $(PA)_2FAPb_2I_7$ originates from excitons localised within the periodic potential of the moiré superlattice generated by rotational misalignment between twin-related crystal layers. The characteristic periodicity of ca. 145 meV between quantised levels, the exceptional thermal robustness of $X_5$, and the high polarisation anisotropy of the moiré-confined states all point to a spatially modulated excitonic landscape with properties qualitatively distinct from defects or STEs[17,18,52,53].

The formation and character of a moiré superlattice are governed by whether the two constituent rotated lattices satisfy a commensurability condition (Supplementary Note 14). For two rotated lattices, the commensurability condition[57] is defined by the existence of integer pairs ($m_1$, $m_2$, $n_1$, $n_2$) satisfying

$$R\left(-\frac{\theta}{2}\right)(m_1 a_1^t + m_2 a_2^t) = R\left(\frac{\theta}{2}\right)(n_1 a_1^b + n_2 a_2^b) \quad (2)$$

$$where, \qquad R(\varphi) = \begin{pmatrix} cos\varphi & -sin\varphi \\ sin\varphi & cos\varphi \end{pmatrix}$$

$R\left(\pm\frac{\theta}{2}\right)$ are counter-rotations by half the relative twist angle, and $a_i^{t,b}$ are the primitive lattice vectors of the top (t) and bottom (b) layers, respectively[57]. A systematic search over all physically relevant twist angles for the orthorhombic lattice of $(PA)_2FAPb_2I_7$ identified only four commensurate configurations (θ = 6˚, 10˚, 17˚, and 19˚), in which the atomic periodicity repeats uniformly across the supercell and a well-defined periodic potential landscape is established (Supplementary Table 5). The experimentally observed twist angle of approximately 5.17˚ lies outside these commensurate values, placing the crystal in a quasi-incommensurate stacking geometry with only short-range lateral periodicity under ambient conditions.

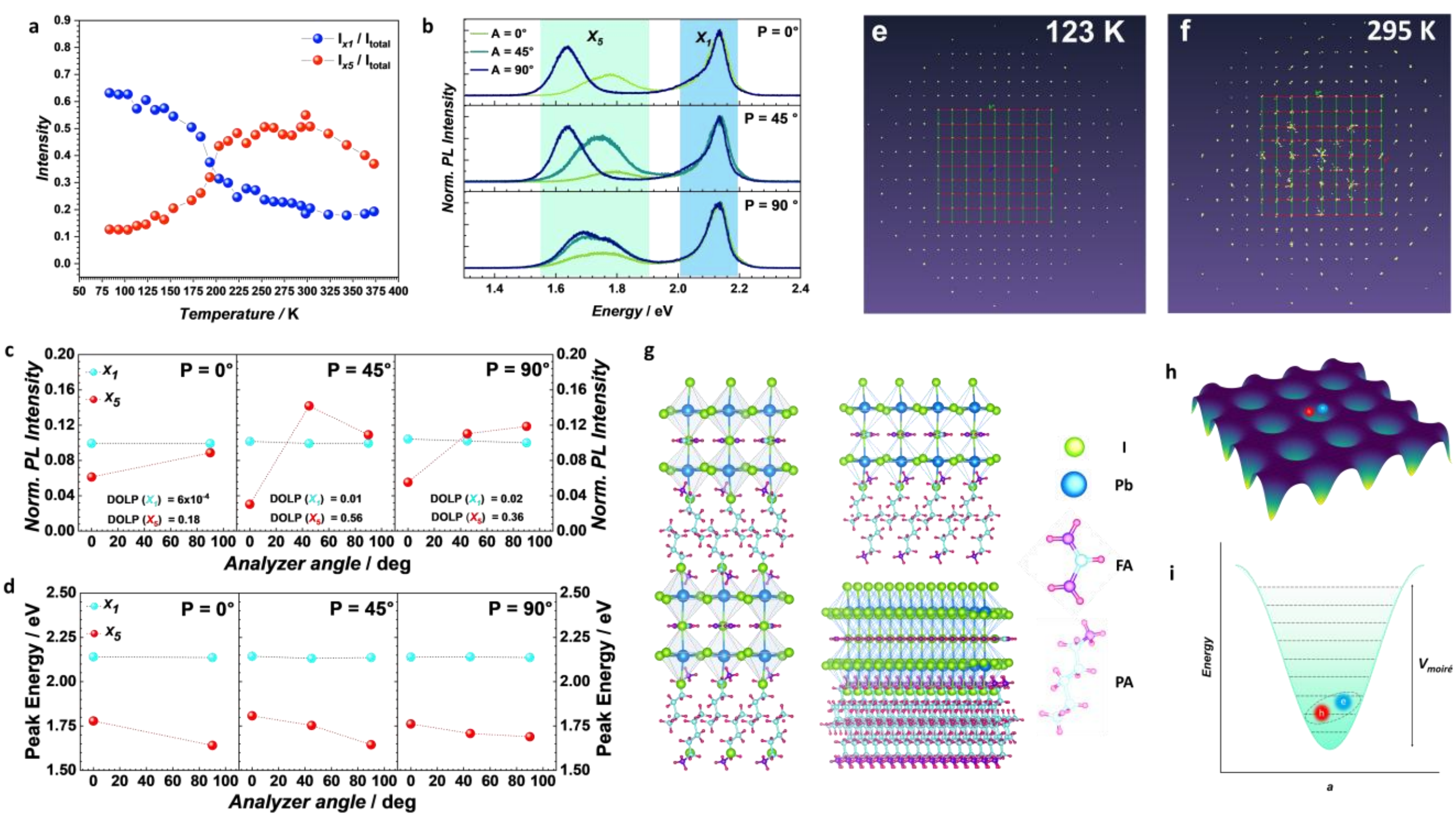

**Figure 5. Moiré exciton formation in twisted layers of 2D $(PA)_2FAPb_2I_7$ perovskite flakes.**
**(a)** Temperature-dependent Integrated PL Intensity for PL peak $X_1$ and $X_5$ of exfoliated single crystal $(PA)_2FAPb_2I_7$. **(b)** Polarisation angle-resolved PL spectra measured at RT for exfoliated single crystal $(PA)_2FAPb_2I_7$ sample with varying analyzer angle. **(c-d)** Polarisation-resolved PL intensity (panel c) and peak energy (panel d) for two emission features, $X_1$ (cyan) and $X_5$ (red), measured as a function of analyzer angle under three different excitation polarisation settings (P = 0, 45, and 90). DOLP values indicate the degree of linear polarisation for each feature. $X_5$ displays pronounced intensity anisotropy and moderate polarisation-dependent energy shifts, while $X_1$ shows nearly constant intensity and peak energy across all polarisation settings. **(e-f)** Reciprocal lattice plots of the $(PA)_2FAPb_2I_7$ single crystals measured at 123 K and 295 K, respectively. **(g)** Non-twisted structure of $(PA)_2FAPb_2I_7$ in Pnma phase (left panel) and Twisted bilayer structure of $(PA)_2FAPb_2I_7$ in Pnma phase (right panel) **(h)** Illustration of the superlattice formed in a twisted homobilayer, which creates a spatially periodic modulation of the electrostatic potential. This periodic potential landscape can confine excitons, localizing them within the minima of the pattern. **(i)** Illustration of how the potential $V_{moiré}$ varies across lattice in a twisted bilayer system, along with the spatial localization of moiré excitons trapped within regions of minimum potential.

Temperature-dependent diffraction directly supports this mechanism. At 123 K, sharp periodic superlattice reflections indicate long-range commensurate order (Figure 5e), whereas at 298 K these reflections broaden and lose coherence, evidencing a commensurate-incommensurate transition (Figure 5f). Correspondingly, the low-temperature commensurate moiré potential localizes excitons, producing the quantized $X_2$-$X_5$ PL ladder, while loss of commensurability at room temperature delocalizes excitons, suppresses miniband quantization, and collapses the spectrum into a single broadened emission (Supplementary Note 13). Taken together, these results establish a direct and quantitative link between moiré commensurability, structural coherence, and the quantised optical response of $(PA)_2FAPb_2I_7$.

## DISCUSSION

This work identifies intrinsically coupled phenomena in 2D hybrid perovskites: (i) $(PA)_2FAPb_2I_7$, exhibiting unambiguous RT piezoelectricity; (ii) polar symmetry arising through a hybrid improper ferroelectric (HIF) mechanism; and (iii) spontaneous moiré superlattice formation originating from pseudo-merohedral twinning, which explains the previously unresolved quantised excitonic emission. All three phenomena arise from a common structural origin: trilinear coupling between the octahedral rotational $X_2^+$ mode, octahedral tilt $X_3^-$ mode, and secondary $\Gamma_4^-$ polar displacements.

The anomalous secondary PL emissions, are conclusively assigned here to moiré-confined excitons through power-, temperature-, and polarisation-resolved spectroscopy combined with theoretical analysis. Concurrently, room-temperature piezoelectricity is demonstrated via PFM with $d_{33}$ (ca. 20 pm/V) without external poling, enabled by the HIF-driven inversion-symmetry breaking. The HIF mechanism, first established in layered oxide perovskites by Benedek and Fennie[58], has only rarely been experimentally realised in halide-based hybrid perovskites[59].

Unlike artificially assembled twisted perovskite bilayers[18,60,61], the ~5.17° twist angle and associated moiré physics emerge spontaneously during solution growth. The thermally driven commensurate-incommensurate transition dynamically switches the moiré potential between ordered and disordered regimes, representing a behaviour absent in conventional vdW moiré systems.

These findings redefine pseudo-merohedral twinning from a crystallographic imperfection into a chemically tuneable route to moiré superlattices accessible through scalable solution processing, consistent with emerging perspectives on perovskite twistronics beyond vdW materials[62]. The resulting moiré-confined excitonic states, exhibiting strong polarisation anisotropy and valley-selective selection rules, provide a potential platform for programmable quantum emitter arrays and spin-orbit-coupled artificial lattices[63].

Technologically, the RT $d_{33}$ (ca. 20 pm/V) places $(PA)_2FAPb_2I_7$ competitively within reported quasi-2D perovskite piezoelectric families[64], while its visible-range bandgap overcomes limitations associated with wide-bandgap ferroelectric perovskites[65,66]. The non-centrosymmetric *Pna2₁* structure additionally supports switchable polarisation behaviour analogous to recent 2D RP ferroelectrics[67], whereas the moiré excitons offer a chemically tuneable platform for quantum-photonic functionality comparable to emerging moiré cavity-QED and superlattice quantum-device architectures[68,69]. Collectively, these results position structural complexity in $(PA)_2FAPb_2I_7$ not as a limitation, but as a multifunctional asset enabling coupled photoferroelectrics, electro-optic, and quantum-photonic applications.

# METHODS

## Materials

All starting materials for synthesis were purchased commercially and were used without further purification. Lead(II) oxide powder, (< 10 $\mu$m), ReagentPlus®, (≥ 99.9%) trace metals basis, Hypophosphorous acid solution 50 wt. % in $H_2O$ and Hydroiodic acid 57 wt. % in $H_2O$, distilled, stabilized, 99.95%, were purchased from Sigma Aldrich. Pentylammonium hydrochloride (≥ 98%) and formamidinium hydrochloride (≥ 98%) were purchased from GreatCell Solar. Analytical grade Acetone, Isopropyl alcohol (IPA), and Ethanol were used. Deuterated dimethyl sulfoxide (DMSO-$d_6$) was used for NMR. The same batch of the starting materials were used in all syntheses.

## Synthesis of Single Crystal flakes

$(PA)_2FAPb_2I_7$ Single crystal flakes were synthesized adapting the temperature-lowering method[70] via reflux reaction setup. Lead(II) oxide (2 mmol) was dissolved in 57% w/w aqueous HI solution (24.4 mmol) in a round bottom flask by heating to 230 °C under constant magnetic stirring, which formed a bright yellow solution. Formamidinium chloride (1 mmol) and n-pentylamonium chloride (1.38 mmol) in 50% w/w aqueous $H_3PO_2$ (3.8 mmol) solution was prepared separately in another vial. Subsequently, upon drop-wise addition of organic cation mixture to the Pb-based solution, few red crystallites were formed, which dissolve rapidly. The combined clear solution was stirred for another 10 min at 230 °C. The stirring was then discontinued, and the solution was cooled down using step-cooling method to avoid crystallization of any unwanted secondary phases. The solution was first cooled to 127 °C, at which point few red flat crystals began to precipitate. The solution temperature was maintained at this temperature to ensure uniform growth of crystals. The solution was then slowly cooled to 50 °C (with a rate of 3 °C per 10 min). The crystals were isolated by suction filtration and thoroughly dried in a vacuum oven at 65 °C overnight.

## Single Crystal Exfoliation

Prior to exfoliation, all the substrates (Si wafers, ITO Glass, Glass cover slips) were cleaned in acetone and IPA solvent by ultrasonication treatment for 10 min each. $(PA)_2FAPb_2I_7$ crystal flakes were mechanically exfoliated using the Scotch tape, followed by PDMS (gel film) tape. The as-cleaved surface of the crystal on PDMS (gel film) tape was positioned in direct contact with the substrate, then gently pressed to ensure uniform exfoliation of crystals and instantly removed from the substrate. The as-obtained nanometer-sized delaminated Bulk and atomically thin layers of single crystal were identified through an optical microscope and used for further characterizations.

## X-Ray Diffraction (XRD)

**Powder X-Ray Diffraction (PXRD)** X-ray diffraction (XRD) data was collected using a Bruker D8 ADVANCE diffractometer equipped with Cu-$k_\alpha$ ($\lambda$ = 0.15406 nm), generated from a 1600 W Cu anode and filtered through Ni filters. Measurements were performed under ambient conditions using an Anton Paar HTK 1200N chamber, with the sample rotating at 15 rpm to improve signal averaging. The scan was carried out in continuous mode over a $2\theta$ range of 2 ° to 40 °, using a step size of 0.02 ° and a dwell time of 1.2 seconds per step.

**Variable Temperature high resolution Powder X-ray diffraction (PXRD)** The XRD data was collected using the Rigaku SmartLab X-ray diffractometer with Cu-$K_\alpha$ radiation ($\lambda$ = 0.15406 nm). The Cu rotating anode X-ray source was operated at a voltage of 45 kV and current 200 mA. The high intensity incident divergent beam was achieved by the CBO-$\alpha$ optics and a 2.5° Soller slit without using the Ni-filter. The dynamic divergence slit (with variable slit width) was used to maintain a 15 mm irradiation width over the sample. An anti-scattering plate was positioned before the receiving 2.5° Soller slit for a low scattering background. The in-situ XRD measurements were conducted at various temperatures under vacuum using the Anton Paar DSC500 domed stage. The XRD measurements started from 25 °C with a vacuum level of 5 x $10^{-2}$ to 9 x $10^{-2}$ mbar.

**Single Crystal X-Ray Diffraction (SCXRD)** A representative red-orange thin plate crystal with dimensions 0.144x0.131x0.021 mm was selected and mounted on a nylon cryoloop. Diffraction data were collected at 123 K using Mo-$K_\alpha$ radiation (l = 0.71073 Å) on a Rigaku Synergy S diffractometer fitted with a HYPIX 6000 hybrid photon counting detector. Data were collected and processed, including a gaussian absorption correction, with the CrysAlisPro software. The structure was solved and refined by standard methods using the SHELX software suite in conjunction with the Olex2 graphical interface. Non-hydrogen atoms were refined with anisotropic displacement ellipsoids and hydrogen atoms were placed in calculated positions using a riding model. The formamidinium cation was refined with restrained anisotropic displacement parameter (RIGU)., The pentylammonium cation was modelled as disordered with two positions each for $C_2$ and $C_5$, corresponding to different configurations of the pentyl chain, with refined occupancies of 0.71:0.29. The geometry and anisotropic displacement of the disorder components was restrained (SADI, SIMU, RIGU, ISOR). For the final refinement cycles, the data were modelled as a pseudo merohedral twin (TWIN 0 0 1 0 1 0 -1 0 0, BASF 0.18). A second sample with dimensions 0.158x0.138x0.026 mm was measured at room temperature, and the structure was refined using the above solution as a starting model. Only the heavy atoms, I, Pb were refined anisotropically.

## Scanning Electron Microscope (SEM)

Scanning electron microscopic coupled to energy dispersive X-ray spectroscopic analysis was performed using a field emission gun scanning electron microscope (Verios 5UC FEGSEM) equipped with a Bruker Quantax 400 X-ray analysis system and 60 $mm^2$ SDD, retractable CBS/DBS BSE detector, and a Python-based application programming interface.

## Nuclear magnetic resonance (NMR) Spectroscopy

The NMR sample was prepared by dissolving *ca.* 1 mg of single crystal flakes in 1 mL of DMSO-$d_6$ single dose (Cambridge Isotope laboratories, Inc.). The NMR tubes were stored in an oven at 80 °C. NMR spectroscopic analysis was performed using a Bruker Avance Neo Nanobay spectrometer (400.20 MHz, $^1$H; 100.6 MHz, $^{13}$C) with a BBFO probe at 25 °C. NMR spectra were processed using Bruker TOPSPIN 4.1.4 software. $^1$H and $^{13}$C spectra were recorded using a sweep width of 8 and 20 kHz respectively, and a prescan delay of 6.5 $\mu$s for both. The acquisition time (AQ) was 4 s, $d_1$ was 1 s, and the total number of scans was 16 for $^1$H. While the AQ was 1.4 s, $d_1$ was 2 s and the total number of scans was 1024 for $^{13}$C.

## Differential Scanning Calorimetry (DSC)

DSC measurements were performed on a PerkinElmer 8000 differential scanning calorimeter. The equipment was calibrated using indium ($T_m$ = 156.6 °C, $\Delta H_m$ = 28.45J/g). Measurements were performed under a nitrogen atmosphere ($N_2$ flow rate of 20ml/min). An amount of *ca.* 2.5 mg sample was sealed in an aluminum pan, with heating and cooling rate of 10 °C/min and 1 min isothermal at both temperature limits in triplicate cycles. Thermal data were recorded from the second heating–cooling cycle.

## Piezo Force Microscopy (PFM)

PFM imaging and measurements were performed on Dimension Icon Microscope with Nanoscope VI controller from Bruker Technologies in a class 10000 clean room setup. SCM-PIT-V2 (Bruker) Platinum-Iridium coated tip with a spring constant of 3 N/m and tip end radius of 25 nm was used for all PFM measurements. AC bias was applied through the tip, and induced sample deformation whose amplitude and phase represent the magnitude of the piezoelectric coefficient and the polarisation direction of the response, was detected respectively. It was driven under an AC bias voltage in the range of $V_{AC}$ = 2-7V and frequencies were around 230-270 KHz. In order to avoid mixing of mechanical response with piezoelectric response, measurements were performed in off-resonance mode. To avoid drifting of the contact resonance frequency due to mechanical changes, contact resonance frequency sweeps were performed before every scan to recalibrate to sub-contact resonance frequency. For $d_{33}$ quantification, a relatively small and optimal drive voltages were used to achieve a good signal-noise ratio and to avoid polarisation switching and non-linearity effect.

## Temperature-dependent PL Spectroscopy

Photoluminescence measurements were conducted using Confocal Microscope System (Witec Alpha 300R) with a 100x objective lens (NA = 0.9) in ambient condition. A fiber coupled 532 nm CW laser having a spot size of 1 $\mu$m diameter was used as excitation source. The samples were illuminated from the top side on a piezo-crystal-controlled scanning stage. For temperature-dependent measurements, the sample was placed on a microscope compatible Linkam chamber connected to a temperature controller; liquid nitrogen was used as the coolant.

## Inductively coupled plasma mass spectrometry (ICP-MS)

Due to the instability of iodine in the $HNO_3$ matrix, two distinct digestion protocols were implemented for the extraction of lead and iodine from $(PA)_2FAPb_2I_7$ single crystal. For Pb quantification, a 6.8 mg aliquot of powdered sample was digested overnight in 7 mL of 10% $HNO_3$ (*ca.* 1.5 M). The resulting solutions underwent two sequential 100-fold dilutions with 2 wt.% HNO3 (*ca.* 0.3 M) to obtain a total dilution factor of 10,000. Instrumental drift and matrix effects were corrected by introducing $^{115}$In as internal standard. Calibration standards were prepared by diluting a commercial Pb stock solution (100 ppm, PerkinElmer) with 0.3 M $HNO_3$, with triplicate measurements used to establish signal-concentration relationships. (Figure S1-a) For Iodine quantification, a 3.5 mg portion of the same powder was dissolved in 4 mL of 5 M KOH, followed by two sequential 100-fold dilutions with ultrapure water to obtain a total dilution factor of 10,000. $^{34}$S (50 ppb) in ultrapure water served as the internal standard. Iodine calibration curve (Figure S1-b) was generated using a $^{127}$I stock solution (1000 ppm, Choice Analytical, Australia) diluted with ultrapure water, with triplicate analyses performed for each standard. All measurements were conducted using a PerkinElmer NexION 2000 ICP-MS. Triplicate analyses showed relative standard deviations <3%. Background corrections were applied using matrix-matched blanks: 2 wt.% $HNO_3$ for Pb determinations and ultrapure water for iodine analyses.

## High Resolution Transmission Electron Microscopy (TEM)

Transmission electron microscopy (TEM) was performed using a FEI Tecnai F20 FEGTEM operated at an accelerating voltage of 200 kV. Samples were dispersed in ethanol via 10 minutes of ultrasonication at room temperature, then drop-cast onto a lacey carbon copper grid and air-dried before imaging.

## Computational Details

The first-principles density functional theory (DFT) and density functional perturbation theory (DFPT) calculations were performed using the Vienna ab initio simulation package (VASP).[88] These simulations were performed using the Perdew-Burke-Ernzerhof (PBE) exchange-correlation functional[71] and projector-augmented wave (PAW) pseudopotentials[72] with a 520 eV energy cutoff. To account for van der Waals interactions, Becke-Johnson (BJ) damping variant of DFT-D3[73] was implemented. The atomic structures of the twisted $(PA)_2FAPb_2I_7$ system were generated using the CellMatch[74] software. Only the $\Gamma$ point was sampled in the Brillouin zone to obtain the piezoelectric tensor.

Phonon properties were computed using the finite-displacement method[75] as implemented in Phonopy. Supercells were constructed to capture interatomic force constants, and atomic displacements were generated symmetrically. The forces obtained from VASP were used to extract harmonic force constants. To accurately describe long-range Coulomb interactions in this polar system, non-analytical corrections (NAC)[76] were included using Born effective charges and dielectric tensors. To incorporate higher-order anharmonic interactions and improve the description of lattice dynamics, all phonon dispersions, irreducible representations and mode analyses were performed using the combined Phonopy-HiPhive[77] workflow, ensuring a consistent description of lattice instabilities and symmetry-adapted distortion modes.

## AUTHOR CONTRIBUTION

AA, JJJ and ANS conceived and supervised the project. SSP was primarily responsible for conceptualizing the study, conducting the literature survey, identifying the research gap, and drafting the manuscript. SSP synthesized the 2D perovskite single crystals, assisted in sample preparation for characterizations, coordinated the overall experimental activities, assisted in analysis and drafting of the PXRD, temperature-variable PXRD, SCXRD, PL, and HRTEM. SSP also performed the DSC measurements, carried out all DFT simulations and theoretical calculations, and led the overall data analysis, interpretation, and figure preparation for the manuscript. SR performed and analysed the AFM and PFM measurements; and also performed all PL measurements, and contributed to the figure preparation and manuscript drafting of the piezoelectric properties section. MB was involved in synthesis and coordinated the overall experimental activities with SSP, and performed and analysed the SCXRD and NMR measurements. MB also provided critical feedback and revised the manuscript. NN performed the PXRD measurements and assisted in the HRTEM analysis. CF and AM performed and analysed the SCXRD measurements. TN led the high-resolution TEM characterization. TVM performed and analysed the ICP-MS measurements. LN performed and analysed the SEM measurements. NM performed the variable temperature high resolution PXRD measurements. ANS, AA, and JJJ reviewed, revised, and refined the manuscript. All authors discussed the results and approved the final version of the manuscript.

## ACKNOWLEDGEMENTS

Authors acknowledge: (i) Dr. Anders Barlow and University of Melbourne for assistance with the Raman measurements, which are not included in the present work since excessive fluorescence impeded reliable analysis with no conclusive results; (ii) the use of PARAM Rudra and Praganak High-Performance Computing (HPC) facility at IIT Bombay, provided under the National Supercomputing Mission (NSM), Government of India; (iii) the computational support by MASSIVE HPC facility (www.massive.org.au) and the Monash eResearch Centre and eSolutions-Research Support Services through the use of the MonARCH HPC Cluster. A part of this research was undertaken with the assistance of resources from the National Computational Infrastructure (NCI Australia), an NCRIS enabled capability supported by the Australian Government. This work was funded by the Australian Research Council (Future Fellowship to ANS; FT200100317), and the IITB-Monash PhD program (fellowship to SSP under the supervision of AA, ANS and JJ).

# REFERENCES

1. Cheng, B. & others. Extremely reduced dielectric confinement in two-dimensional hybrid perovskites with large polar organics. *Commun. Phys.* **1**, 80 (2018).

2. Gluchowski, A. & others. Polarons and exciton polarons in two-dimensional polar materials. *Phys. Rev. B* **112**, 45413 (2025).

3. Guo, S. & others. Exciton engineering of 2D Ruddlesden–Popper perovskites. *Nat. Commun.* **15**, 2922 (2024).

4. Tao, W., Zhang, Y. & Zhu, H. Dynamic exciton polaron in two-dimensional lead halide perovskites and implications for optoelectronic applications. *Acc. Chem. Res.* **55**, 345–353 (2022).

5. Tao, W., Zhang, C., Zhou, Q., Zhao, Y. & Zhu, H. Momentarily trapped exciton polaron in two-dimensional lead halide perovskites. *Nat. Commun.* **12**, 1400 (2021).

6. Lin, M.-L. & others. Correlating symmetries of low-frequency vibrations and self-trapped excitons in layered perovskites for light emission with different colors. *Small* **18**, 2106759 (2022).

7. Krahne, R., Lin, M.-L. & Tan, P.-H. Interplay of phonon directionality and emission polarization in two-dimensional layered metal halide perovskites. *Acc. Chem. Res.* **57**, 2476–2489 (2024).

8. Castelli, A. & others. Revealing photoluminescence modulation from layered halide perovskite microcrystals upon cyclic compression. *Advanced Materials* **31**, 1805608 (2019).

9. Blancon, J.-C. & others. Extremely efficient internal exciton dissociation through edge states in layered 2D perovskites. *Science (1979).* **355**, 1288–1292 (2017).

10. Kinigstein, E. D. & others. Edge states drive exciton dissociation in Ruddlesden–Popper lead halide perovskite thin films. *ACS Mater. Lett.* **2**, 1360–1367 (2020).

11. Qin, Z. & others. Spontaneous formation of 2D/3D heterostructures on the edges of 2D Ruddlesden-Popper hybrid perovskite crystals. *Chemistry of Materials* **32**, 5009–5015 (2020).

12. Shi, E. & others. Extrinsic and dynamic edge states of two-dimensional lead halide perovskites. *ACS Nano* **13**, 1635–1644 (2019).

13. Hong, J., Prendergast, D. & Tan, L. Z. Layer edge states stabilized by internal electric fields in two-dimensional hybrid perovskites. *Nano Lett.* **21**, 182–188 (2020).

14. Qin, Y. & others. Dangling octahedra enable edge states in 2D lead halide perovskites. *Advanced Materials* **34**, 2201666 (2022).

15. Jin, C. & others. Observation of moiré excitons in WSe2/WS2 heterostructure superlattices. *Nature* **567**, 76–80 (2019).

16. Das, P. & Chattopadhyay, A. Moiré superlattices of two-dimensional copper nanocluster assemblies with tuneable twin emissions from hierarchical components leading to white light emission. *J. Mater. Chem. C Mater.* **11**, 12029–12036 (2023).

17. Zhang, S. & others. Moiré superlattices in twisted two-dimensional halide perovskites. *Nat. Mater.* **23**, 1222–1229 (2024).

18. Zhang, L., Zhang, X. & Lu, G. Predictions of moiré excitons in twisted two-dimensional organic–inorganic halide perovskites. *Chem. Sci.* **12**, 6073–6080 (2021).

19. Lu, J. & others. Origin and physical effects of edge states in two-dimensional Ruddlesden-Popper perovskites. *iScience* **25**, (2022).

20. Siwach, P., Sikarwar, P., Halpati, J. S. & Chandiran, A. K. Design of above-room-temperature ferroelectric two-dimensional layered halide perovskites. *J. Mater. Chem. A Mater.* **10**, 8719–8738 (2022).

21. Rong, S.-S., Faheem, M. B. & Li, Y.-B. Perovskite single crystals: Synthesis, properties, and applications. *Journal of Electronic Science and Technology* **19**, 100081 (2021).

22. Le Bail, A., Duroy, H. & Fourquet, J. L. Ab-initio structure determination of LiSbWO6 by x-ray powder diffraction. *Mater. Res. Bull.* **23**, 447–452 (1988).

23. Oh, Y. S. & others. Experimental demonstration of hybrid improper ferroelectricity and the presence of abundant charged walls in (Ca,Sr)3Ti2O7 crystals. *Nat. Mater.* **14**, 407–413 (2015).

24. Smith, K. A., Nowadnick, E. A., Fan, S. & others. Infrared nano-spectroscopy of ferroelastic domain walls in hybrid improper ferroelectric Ca3Ti2O7. *Nat. Commun.* **10**, 5235 (2019).

25. Sung, S. H. & others. Torsional periodic lattice distortions and diffraction of twisted 2D materials. *Nat. Commun.* **13**, 7826 (2022).

26. Cao, Y. & others. Unconventional superconductivity in magic-angle graphene superlattices. *Nature* **556**, 43–50 (2018).

27. Pica, G. & others. Photo-ferroelectric perovskite interfaces for boosting VOC in efficient perovskite solar cells. *Nat. Commun.* **15**, 8753 (2024).

28. Sharma, P. & others. A room-temperature ferroelectric semimetal. *Sci. Adv.* **5**, eaax5080 (2019).

29. Seol, D., Kim, B. & Kim, Y. Non-piezoelectric effects in piezoresponse force microscopy. *Current Applied Physics* **17**, 661–674 (2017).

30. Xue, F. & others. Multidirection piezoelectricity in mono- and multilayered hexagonal α-In2Se3. *ACS Nano* **12**, 4976–4983 (2018).

31. Garduño-Medina, A., Vázquez-Delgado, M., Diliegros-Godines, C., García-Vázquez, V. & Flores-Ruiz, F. Measurements outside resonance with piezoresponse force microscopy. in *AIP Conference Proceedings* vol. 2416 (AIP Publishing, 2021).

32. Gomez, A., Puig, T. & Obradors, X. Diminish electrostatic in piezoresponse force microscopy through longer or ultra-stiff tips. *Appl. Surf. Sci.* **439**, 577–582 (2018).

33. Killgore, J. P., Robins, L. & Collins, L. Electrostatically-blind quantitative piezoresponse force microscopy free of distributed-force artifacts. *Nanoscale Adv.* **4**, 2036–2045 (2022).

34. Kim, S., Seol, D., Lu, X., Alexe, M. & Kim, Y. Electrostatic-free piezoresponse force microscopy. *Sci. Rep.* **7**, 41657 (2017).

35. Sezer, N. & Koç, M. A comprehensive review on the state-of-the-art of piezoelectric energy harvesting. *Nano Energy* **80**, 105567 (2021).

36. Lee, B. Y. & others. Virus-based piezoelectric energy generation. *Nat. Nanotechnol.* **7**, 351–356 (2012).

37. Kim, D. M. & others. Thickness dependence of structural and piezoelectric properties of epitaxial Pb(Zr0.52Ti0.48)O3 films on Si and SrTiO3 substrates. *Appl. Phys. Lett.* **88**, (2006).

38. Rahmany, S. & others. The impact of piezoelectricity in low dimensional metal halide perovskite. *ACS Energy Lett.* **9**, 1527–1536 (2024).

39. Hautzinger, M. P., Mihalyi-Koch, W. & Jin, S. A-site cation chemistry in halide perovskites. *Chemistry of Materials* **36**, 10408–10420 (2024).

40. Frost, J. M. & others. Atomistic origins of high-performance in hybrid halide perovskite solar cells. *Nano Lett.* **14**, 2584–2590 (2014).

41. Li, C. & others. Formability of ABX3 (X = F, Cl, Br, I) halide perovskites. *Acta Crystallogr. B* **64**, 702–707 (2008).

42. Feng, Y. & others. Thickness-dependent evolution of piezoresponses and a/c domains in [101]-oriented PbTiO3 ferroelectric films. *J. Appl. Phys.* **128**, (2020).

43. Feng, Y. & others. Thickness-dependent evolution of piezoresponses and stripe 90 domains in (101)-oriented ferroelectric PbTiO3 thin films. *ACS Appl. Mater. Interfaces* **10**, 24627–24637 (2018).

44. Kelley, K. P. & others. Thickness and strain dependence of piezoelectric coefficient in BaTiO3 thin films. *Phys. Rev. Mater.* **4**, 24407 (2020).

45. Posmyk, K. & others. Quantification of exciton fine structure splitting in a two-dimensional perovskite compound. *J. Phys. Chem. Lett.* **13**, 4463–4469 (2022).

46. Canet-Albiach, R. & others. Revealing giant exciton fine-structure splitting in two-dimensional perovskites using van der Waals passivation. *Nano Lett.* **22**, 7621–7627 (2022).

47. Dyksik, M. & others. Steric engineering of exciton fine structure in 2D perovskites. *Adv. Energy Mater.* **15**, 2404769 (2025).

48. Wright, A. D. & others. Electron-phonon coupling in hybrid lead halide perovskites. *Nat. Commun.* **7**, 11755 (2016).

49. Paritmongkol, W., Powers, E. R., Dahod, N. S. & Tisdale, W. A. Two origins of broadband emission in multilayered 2D lead iodide perovskites. *J. Phys. Chem. Lett.* **11**, 8565–8572 (2020).

50. Ni, L. & others. Real-time observation of exciton–phonon coupling dynamics in self-assembled hybrid perovskite quantum wells. *ACS Nano* **11**, 10834–10843 (2017).

51. Tongay, S. & others. Defects activated photoluminescence in two-dimensional semiconductors: Interplay between bound, charged and free excitons. *Sci. Rep.* **3**, 2657 (2013).

52. Zhang, L., Zhang, Z., Wu, F. & others. Twist-angle dependence of moiré excitons in WS2/MoSe2 heterobilayers. *Nat. Commun.* **11**, 5888 (2020).

53. Tran, K. & others. Evidence for moiré excitons in van der Waals heterostructures. *Nature* **567**, 71–75 (2019).

54. Jin, F. & others. Exciton polariton condensation in a perovskite moiré flat band at room temperature. *Sci. Adv.* **11**, eadx2361 (2025).

55. Xiao, S. & others. Polarization-dependent multiphoton-excited self-trapped emission in alloyed 0D Rb7Bi3Cl16 metal halides via Sb3+ doping. *Small Science* 2500261 (2025).

56. Förg, M. & others. Moiré excitons in MoSe2-WSe2 heterobilayers and heterotrilayers. *Nat. Commun.* **12**, 1656 (2021).

57. He, Z. & Weng, H. Giant nonlinear Hall effect in twisted bilayer WTe2. *NPJ Quantum Mater.* **6**, 101 (2021).

58. Benedek, N. A. & Fennie, C. J. Hybrid improper ferroelectricity: a mechanism for controllable polarization–magnetization coupling. *Phys. Rev. Lett.* **106**, 107204 (2011).

59. Mao, L. et al. The role of organic cation in hybrid improper ferroelectricity. *J. Am. Chem. Soc.* **143**, 3098–3107 (2021).

60. Ma, X. et al. Square moiré superlattices in twisted two-dimensional halide perovskites. *Nature* **626**, 285–291 (2024).

61. Shree, S. et al. Predictions of moiré excitons in twisted two-dimensional organic–inorganic halide perovskites. *NPJ Comput. Mater.* **7**, 54 (2021).

62. Huang, S. et al. A new platform for twistronics: perovskite moiré superlattices beyond van der Waals. *Small Struct.* https://doi.org/10.1002/sstr.202500770 (2026) doi:10.1002/sstr.202500770.

63. Yu, H., Liu, G.-B., Tang, J., Xu, X. & Yao, W. Moiré excitons: From programmable quantum emitter arrays to spin-orbit–coupled artificial lattices. *Sci. Adv.* **3**, e1701696 (2017).

64. Siol, S. et al. The impact of piezoelectricity in low-dimensional metal halide perovskites. *ACS Energy Lett.* **9**, 1839–1847 (2024).

65. Chen, X.-G. et al. Two-dimensional layered perovskite ferroelectric with giant piezoelectric voltage coefficient. *J. Am. Chem. Soc.* **142**, 1077–1082 (2020).

66. Zheng, W. et al. Emerging halide perovskite ferroelectrics. *Advanced Materials* **35**, 2205410 (2023).

67. Fu, Y. et al. Electrically switchable chiral second-harmonic generation in achiral ferroelectric 2D Ruddlesden–Popper perovskite. *Sci. Adv.* **10**, eadq5521 (2024).

68. Nowakowski, K. et al. Single-photon detection enabled by negative differential conductivity in moiré superlattices. Preprint at (2025).

69. Liang, J. et al. Moiré cavity quantum electrodynamics. *Sci. Adv.* **11**, eadu8732 (2025).

70. Hou, J. & others. Synthesis of 2D perovskite crystals via progressive transformation of quantum well thickness. *Nature Synthesis* **3**, 265–275 (2024).

71. Perdew, J. P., Burke, K. & Ernzerhof, M. Generalized gradient approximation made simple. *Phys. Rev. Lett.* **77**, 3865 (1996).

72. Blöchl, P. E. Projector augmented-wave method. *Phys. Rev. B* **50**, 17953 (1994).

73. Grimme, S., Antony, J., Ehrlich, S. & Krieg, H. A consistent and accurate ab initio parametrization of density functional dispersion correction (DFT-D) for the 94 elements H-Pu. *J. Chem. Phys.* **132**, (2010).

74. Lazić, P. CellMatch: Combining two unit cells into a common supercell with minimal strain. *Comput. Phys. Commun.* **197**, 324–334 (2015).

75. Togo, A. & Tanaka, I. First principles phonon calculations in materials science. *Scr. Mater.* **108**, 1–5 (2015).

76. Wang, Y. *et al.* Mixed-space approach for calculation of vibration-induced dipole-dipole interactions. *Journal of Physics: Condensed Matter* **22**, 202201 (2010).

77. Eriksson, F., Hellman, O., Brorsson, J. & Hellman, O. The HiPhive Package for the Extraction of High-Order Force Constants by Machine Learning. *Adv. Theory Simul.* **2**, 1800184 (2019).

# Supplementary Information for "Hybrid Improper Ferroelectricity and Moiré Superlattices-induced Exciton Quantization in Layered 2D Halide Perovskite"

Sanika S. Padelkar[1,2,3,4,5], Sharidya Rahman[2,3 #], Mattia Belotti[1 #], Naufan Nurrosyid[2,3], Craig Forsyth[1], Alasdair Mckay[1], Tam Nguyen[1], Thi Vu Mung[1], Lan Nguyen[2,3], Naeimeh Mozaffari[2,3], Alexandr N. Simonov[1] *, Aftab Alam[4,5] *, Jacek J. Jasieniak[2,3] *

1. School of Chemistry, Monash University, Melbourne, Victoria, 3800, Australia
2. Department of Materials Science and Engineering, Monash University, Melbourne, Victoria, 3800, Australia
3. Centre of Excellence in Exciton Science, Monash University, Melbourne, Victoria, 3800, Australia.
4. Department of Physics, Indian Institute of Technology Bombay, Powai, Mumbai-400076
5. IITB-Monash Research Academy, Indian Institute of Technology Bombay, Powai, Mumbai-400076

# - These authors contributed equally to this work.

E-mail: jacek.jasieniak@monash.edu; alexandr.simonov@monash.edu; aftab@iitb.ac.in

## Inductively Coupled Plasma Mass Spectrometry (ICP-MS)

## Supplementary Note 1

For Pb quantification, a 6.8 mg aliquot of powdered sample was digested overnight in 7 mL of 10% $HNO_3$ (*ca.* 1.5 M). The resulting solutions underwent two sequential 100-fold dilutions with 2 wt.% HNO3 (*ca.* 0.3 M) to obtain a total dilution factor of 10,000. Instrumental drift and matrix effects were corrected by introducing $^{115}$In, as an internal standard. Calibration standards were prepared by diluting a commercial Pb stock solution (100 ppm, PerkinElmer) with 0.3 M $HNO_3$, with triplicate measurements used to establish signal-concentration relationships. (Supplementary Figure 1a). For Iodine quantification, a 3.5 mg portion of the same powder was dissolved in 4 mL of 5 M KOH, followed by two sequential 100-fold dilutions with ultrapure water to obtain a total dilution factor of 10,000. $^{34}$S (50 ppb) in ultrapure water served as the internal standard. Iodine calibration curve (Supplementary Figure 1b) was generated using a $^{127}$I stock solution (1000 ppm, Choice Analytical, Australia) diluted with ultrapure water, with triplicate analyses performed for each standard.

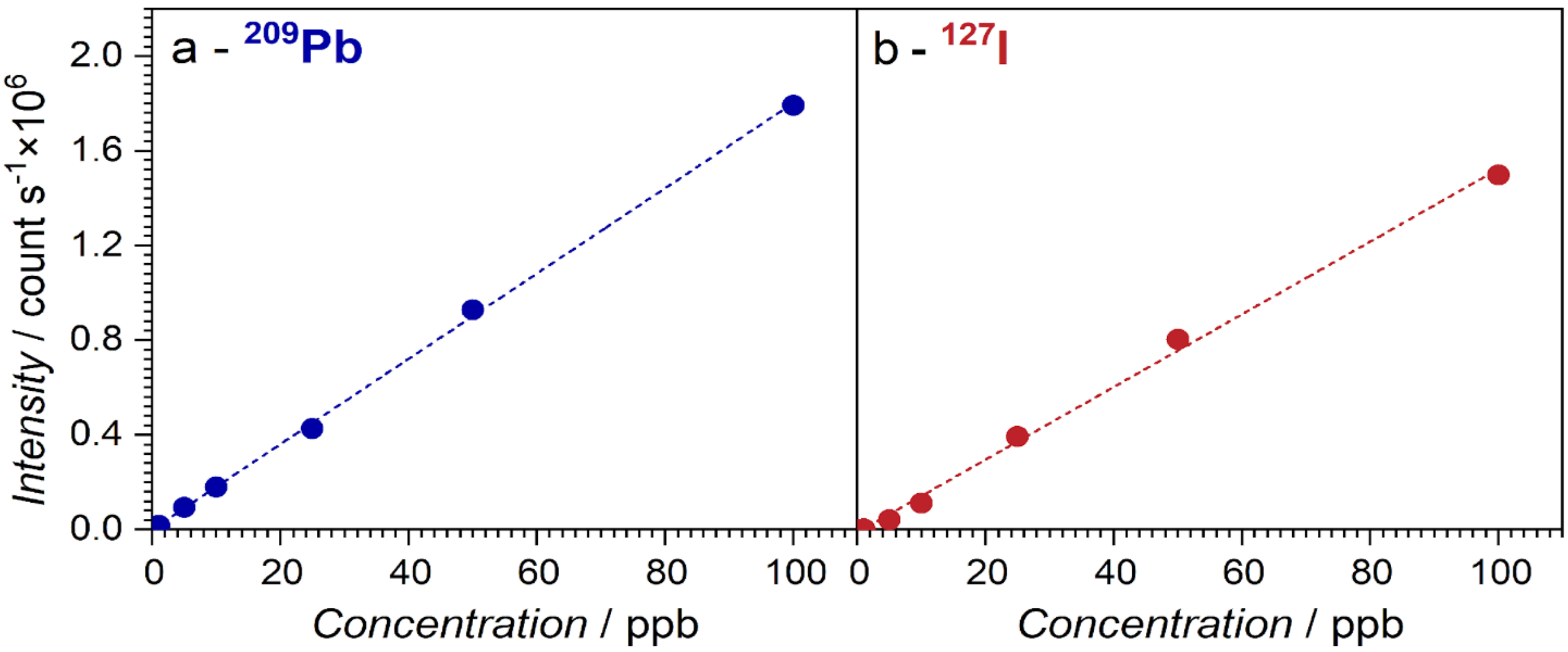

**Figure 1:** ICP-MS calibration plots for **(a)** $^{209}$Pb and **(b)** $^{127}$I Symbols represent experimental data, and dashed lines represent corresponding linear fits. The calibration dependence for figure **(a)** is Intensity [count $s^{-1}$] = (1.801310$^4$ ± 188.6) $C_{Pb}$ [ppb] + (819.7 ± 8205); $R^2$ = 0.9995; and for figure **(b)** is Intensity [count $s^{-1}$] = (1.53510$^4$ ± 352.3) $C_I$ [ppb] + (1.142310$^4$ ± 1.533310$^4$); $R^2$ = 0.9987

## Nuclear magnetic resonance spectroscopy (NMR)

## Supplementary Note 2

**$^1$H-NMR chemical shifts and splitting patterns of PA organic cations in $(PA)_2FAPb_2I_7$ single crystal flakes:**
**6 H** (0.86 ppm, triplet, J = 7.00 Hz): Terminal methyl ($-CH_3-$) protons of the pentyl chain in pentylammonium. The triplet arises from coupling with adjacent methylene protons.

**8 H** (1.27 ppm, multiplet): Two adjacent Methylene (-$CH_2$-) protons in the middle of the pentyl chain, appearing as a multiplet due to coupling with neighboring protons.
**4 H** (1.51 ppm, multiplet): Methylene protons closer to the ammonium group (-$NH_3^+$) in pentylammonium are slightly deshielded.
**4 H** (2.77 ppm, triplet, J = 7.60 Hz): Methylene protons adjacent to the ammonium nitrogen (-$CH_2$-$NH_3^+$), shifted downfield due to the electron-withdrawing effect of the ammonium group.
**6 H** (7.59 ppm, singlet): Ammonium protons (-$NH_3^+$), significantly shifted downfield due to the electron-withdrawing effect of the ammonium group.

**$^1$H-NMR chemical shifts and splitting patterns of FA organic cations in $(PA)_2FAPb_2I_7$ single crystal flakes:**
**1 H** (7.85 ppm, singlet): Methylene (-$CH_2$-) protons in the middle of the formamidinium chain, adjacent to the ammonium group, shifted downfield.
**2 H** (8.64 ppm, singlet): Terminal ammonium proton (-$NH_3^+$), significantly shifted downfield due to the electron-withdrawing effect of the adjacent ammonium group.
**2 H** (8.99 ppm, singlet): Terminal ammonium proton (-$NH_3^+$), significantly shifted downfield due to the electron-withdrawing effect of the adjacent ammonium group.

## Pnma vs $Pna2_1$: Group-Subgroup Relation

## Supplementary Note 3

*Pnma* (No. 62) and *$Pna2_1$* (No. 33) are both orthorhombic space groups belonging to the same primitive lattice, but they differ fundamentally in their point-group symmetry (Supplementary Figure 2). *Pnma* has point group *mmm* ($D_2h$) and is centrosymmetric, while *$Pna2_1$* has point group *mm2* ($C_{2v}$) and is polar. *$Pna2_1$* is the unique maximal translationengleiche subgroup of index 2 of *Pnma*, as tabulated in the International Tables for Crystallography Vol. A1. This means the unit cell dimensions and all lattice translations are identical in both space groups and only half the point-symmetry operations are lost. No intermediate space group exists between them that preserves the same translational lattice, making the relationship mathematically minimal.

The three symmetry elements lost in descending from *Pnma* to *$Pna2_1$* are the inversion centre *i*, the two-fold rotation $C_{2y}$ along b, and the mirror plane $m_z$ perpendicular to b at y = ¼ and ¾. Of these, the loss of $m_z$ is structurally most significant. In *Pnma*, $m_z$ constrains atoms on the 4c Wyckoff site - including the $FA^+$ cation in $(PA)_2FAPb_2I_7$ to a fixed y-coordinate. Its removal in *$Pna2_1$* liberates this coordinate, allowing atomic displacements along the c-axis and generating a unique polar axis along z. The 4c Wyckoff site therefore splits into a general 4a position in *$Pna2_1$* where all three coordinates are free, doubling the asymmetric unit from 1/8 to 1/4 of the unit cell. The loss of inversion symmetry simultaneously activates five independent piezoelectric tensor components ($d_{15}$, $d_{24}$, $d_{31}$, $d_{32}$, $d_{33}$), which are strictly zero in centrosymmetric *Pnma*.

The two space groups are crystallographically nearly indistinguishable by powder XRD for three major reasons. (i) Because the unit cell is preserved, all diffraction peaks appear at identical 2θ positions. (ii) All systematic absences are shared: the n-glide, a-glide, and screw-axis extinction rules are identical in both space groups, so no reflection condition can discriminate between them. (iii) The inversion-breaking atomic displacements are typically of order 0.01-0.1 Å, producing intensity differences below 1-5% of peak intensity below the noise floor of laboratory diffractometers. Pseudo-merohedral twinning further masks the true symmetry: the lost mirror $m_z$ becomes a twin law, and the superimposed diffraction patterns of the two domain variants reconstruct the apparent *Pnma* intensity distribution.

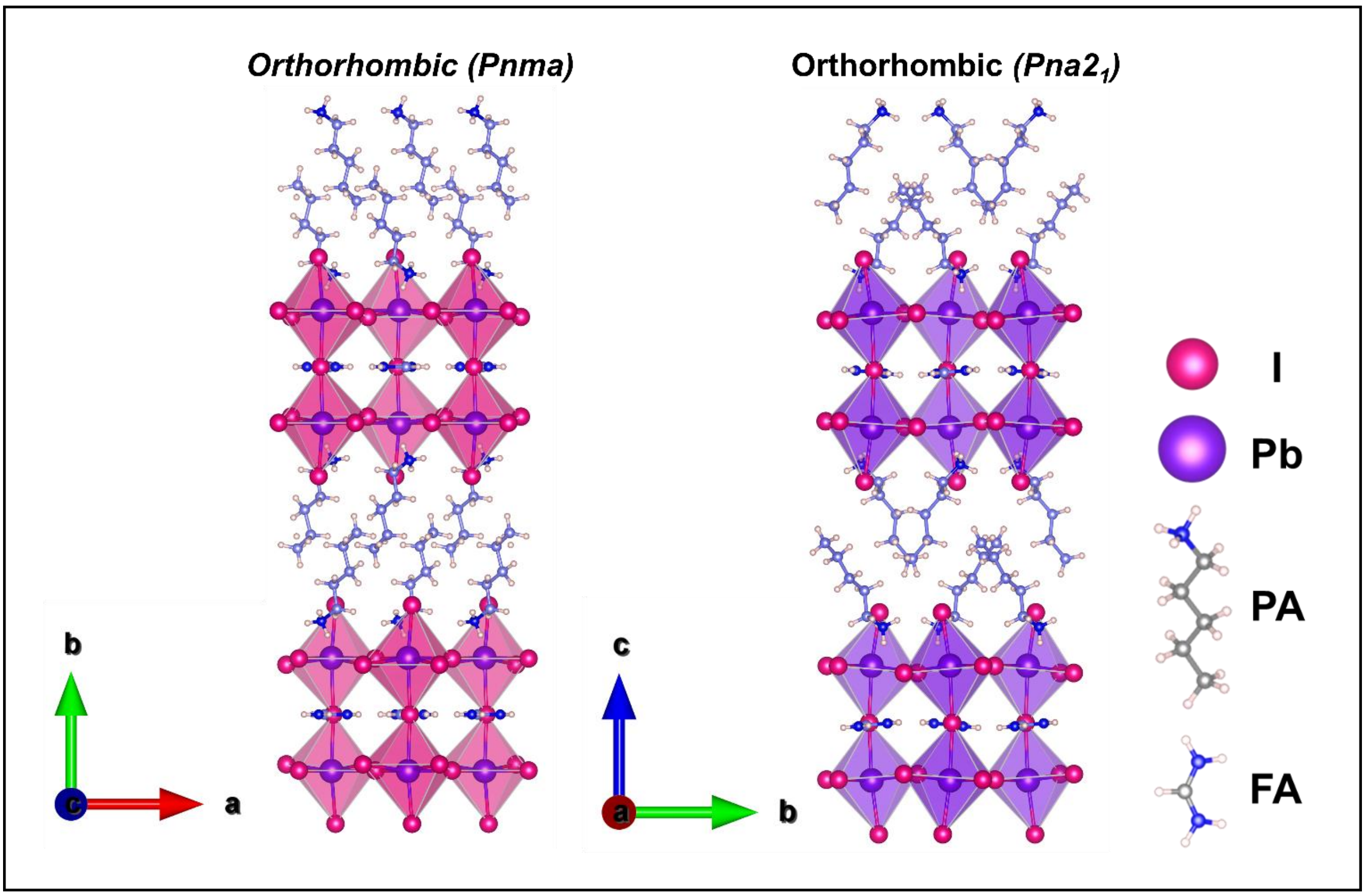

**Figure 2:** Crystal structure of $(PA)_2FAPb_2I_7$ in Pnma and Pna21 space groups derived from the single-crystal XRD analysis.

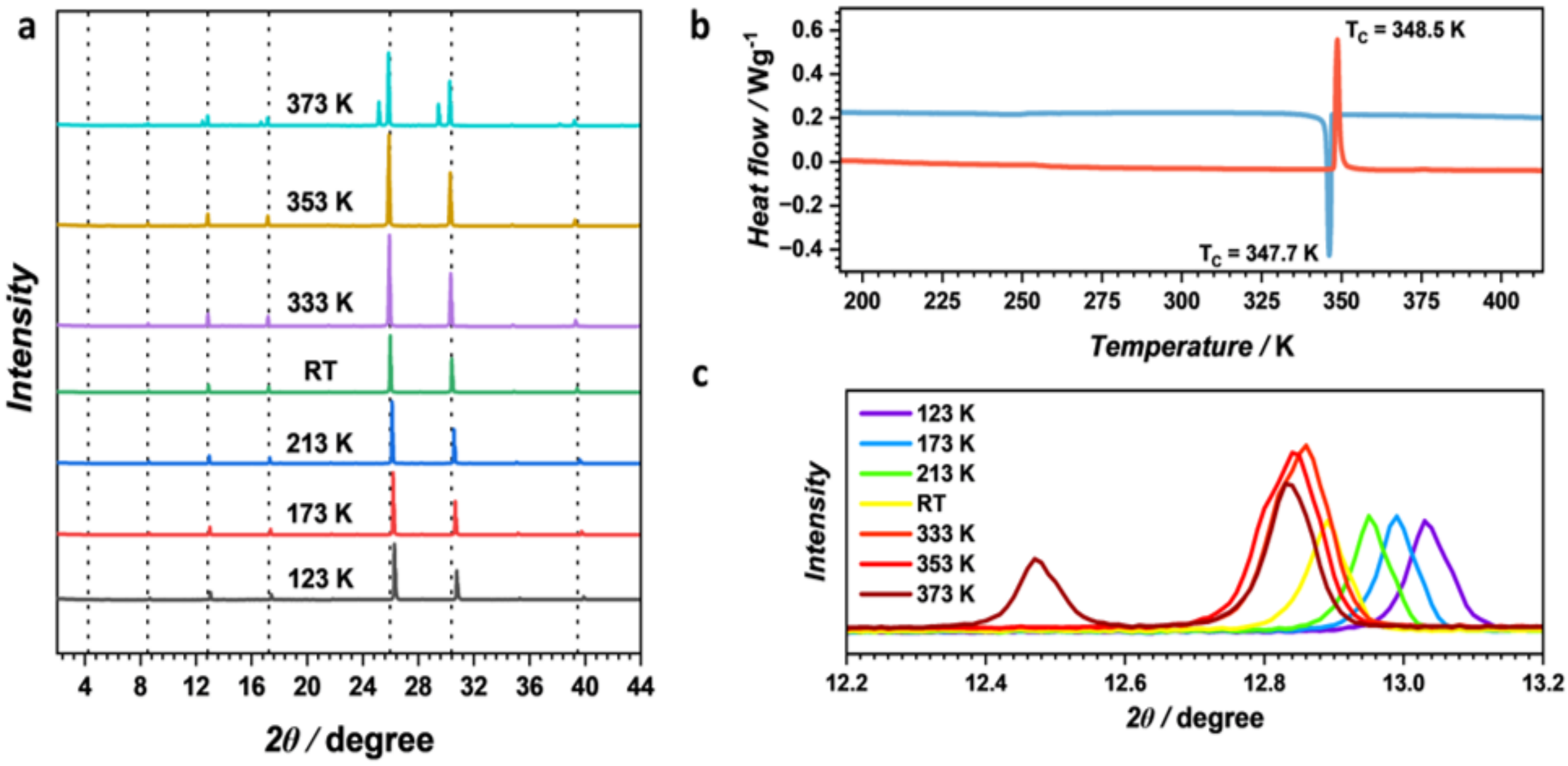

**Figure 3: (a)** Temperature-dependent XRD profile of $(PA)_2FAPb_2I_7$ single crystals from 373 K to 123 K with black dashed lines, emphasizing the peak position at RT **(b)** Differential Scanning Calorimetry (DSC) Curve reveals a phase transition around Curie temperature ($T_c$) ca. 349 K, at which point $(PA)_2FAPb_2I_7$ undergoes a transformation from piezoelectric phase at RT to paraelectric above $T_c$ **(c)** Zoomed-in-view of the XRD pattern from 123 K - 373 K.

## Single Crystal X-Ray Diffraction (SCXRD)

## Supplementary Note 4

The crystal samples of $(PA)_2FAPb_2I_7$ were observed as clusters of small square orange thin plates. These were separated as much as possible and a specimen was mounted onto a nylon cryoloop. In both cases the measured crystal appears as a fragment of a square plate. For the measurements at 123K, each crystal sample was quench cooled prior to data collection. Several samples were measured at each temperature and in all cases, the observed diffraction data indicated the presence of some degree of twinning and/or multi crystal as indicated by the relatively poor fit of the unit cell to the observed reflections (in some cases as low as ca 60 %). For the final data collections, a relatively clean specimen was used for the 123 K data, but the specimen for the 295 K data was less satisfactory (Main Figure 5(e,f)).

Analysis of the diffraction data statistics for sample measured at 123 K, indicated the most likely space groups were orthorhombic *Pnma* (centrosymmetric) or $Pna2_1$ (non-centrosymmetric) based on systematic absences, with a preference for the centrosymmetric space group based on E statistics. The structures were solved in both space groups by heavy atom methods which revealed the positions of the Pb and I atoms. Subsequent refinements revealed plausible positions for the pentylammonium and formamidinium cations in the Difference Fourier map, however the modelling and refinement of the electron density associated with organic moieties required significant restraints on both the geometry and atomic displacement parameters, in particular for the pentylammonium fragment. The difference between the *Pnma* and $Pna2_1$ structures is the presence of a crystallographic mirror plane at y = 0.25/0.75 (running through the formamidine cation and I(4) (Main Figure 2a-b), which means that the crystallographically unique component of the structure (the asymmetric unit) in *Pnma* is half that of the *Pna*$2_1$. For the *Pnma* solution, the pentylammonium cation was modelled as disordered over two positions with refined occupancies 0.71:0.29, representing two differing configurations of the pentyl chain orientation. In the *Pna*$2_1$ refinement, the two configurations occupy separate fully occupied sites. Otherwise, the structures are identical.

**Table 1:** Crystallographic and refinement parameters for $C_{11}H_{33}I_7N_4Pb_2$ at 123K and 295K, in different space groups Pnma and Pna$2_1$

| | **123K** | | **295K** | |
|---|---|---|---|---|
| | $C_{11}H_{33}I_7N_4Pb_2$ | $C_{11}H_{33}I_7N_4Pb_2$ | $C_{11}H_{33}I_7N_4Pb_2$ | $C_{11}H_{33}I_7N_4Pb_2$ |
| | ***Pnma*** | ***Pna2*$_1$** | ***Pnma*** | ***Pna2*$_1$** |
| **Formula** | $C_{11}H_{33}I_7N_4Pb_2$ | | | |
| **M** | 1524.09 | | | |
| **Space group** | Orthorhombic, *Pnma* | Orthorhombic, *Pna*$2_1$ | Orthorhombic, *Pnma* | Orthorhombic, *Pna*$2_1$ |
| $a$ **(Å)** | 8.8590(3) | 8.8590(3) | 8.9492(4) | 8.9492(4) |
| $b$ **(Å)** | 40.2302(13) | 8.9132(3) | 41.083(2) | 9.0171(4) |
| $c$ **(Å)** | 8.9132(3) | 40.2302(13) | 9.0171(4) | 41.083(2) |
| $V$ **(Å$^3$)** | 3176.66(18) | 3176.66(18) | 3315.2(3) | 3315.2(3) |
| $Z$ | 4 | 4 | 4 | 4 |
| $N_t$ | 21698 | 20283 | 24678 | 23068 |
| $N$, $R_{int}$ | 3250, 0.052 | 5772, 0.047 | 3291, 0.095 | 5902, 0.089 |
| $N_0$ ($> 2\sigma$) | 2767 | 4931 | 2365 | 3990 |
| **R1** ($> 2\sigma$) | 0.0547 | 0.0537 | 0.1011 | 0.0846 |
| **wR2 (all data)** | 0.1398 | 0.1502 | 0.1263 | 0.2653 |
| **GoF** | 1.081 | 1.047 | 1.067 | 1.064 |
| $\Delta e_{min,max}$ **(e. Å$^{-3}$)** | −3.17, 3.31 | −5.06, 4.18 | −2.75, 4.21 | −2.79, 4.32 |
| **Flack** | | 0.49(3) | | 0.51(4) |

Analysis of the diffraction data statistics at 295 K showed very similar systematic absences and E statistics to the 123 K data, again indicating either the *Pnma* or *Pna*$2_1$ space groups. The structures were both solved and refined to essentially the same structural models as for the 123 K data. Notably at the higher temperature the atomic displacement parameters are much larger, as expected, and in particular for the organic cations, presumably due to increased thermal motion. The formamidinium cation was most affected and could be modelled as disordered over two positions in *Pna*$2_1$. This is similar to the structural modelling of phase changes observed in some ionic liquid/ plastic crystal structures. However, it should be noted that the residual electron density peaks are significantly high in both the 123 K and 295 K structures although the situation is worse in the 295 K data, presumably reflecting both the high temperature but also the lower quality crystal that was used. The highest of these peaks are observed close to the heavy atom positions and as such may be associated with absorption effects, but there are also other significant peaks that have not been accounted for in both structural models. This means that there is a lot of electron density that is not being modelled by the structure and as such the validity of the model is therefore questionable. Having said that, the calculated and experimental powder patterns do match reasonably well (see below) which does give some support for the current models. In terms of the unit cell data, the 295 K shows an increase in total cell volume of ca 4.4%, with the largest expansion in the 40 Å axis (Supplementary Table 1).

## Lattice Structure and Phonon Dynamics

## Supplementary Note 5

The apparent structural ambiguity in $(PA)_2FAPb_2I_7$ stems from the group–subgroup relationship between *Pnma* and *Pna2₁*. Within a group-theoretical framework based on symmetry-adapted order parameters of the centrosymmetric *Pnma* parent phase, two symmetry-allowed distortion pathways can lead to a non-centrosymmetric $Pna2_1$ structure: (1) zone-center polar displacement mode ($\Gamma_4^-$) at $k = (0,0,0)$, corresponding to proper ferroelectricity, and (2) zone-boundary ($X_2^+$ and $X_3^-$) rotational/tilt mode at $k = (\frac{1}{2}, 0,0)$, which is non-polar by itself, but can lead to improper ferroelectricity. Prior studies on layered inorganic perovskites, such as $Ca_3Ti_2O_7$ and $(Ca,Sr)_3Ti_2O_7$ have demonstrated that such rotational modes can act as primary order parameters, generating ferroelastic twin variants and inducing secondary polarization through multiple distortion mode coupling, a mechanism often referred to as hybrid improper ferroelectricity.

The symmetry reduction from Pnma to $Pna2_1$ is mediated by the $B_{1u}$ irreducible representation of the Pnma factor group at k = 0 (Γ point) (Table 2). $B_{1u}$ carries character −1 under the three lost operations {i, $C_{2y}$, $m_z$} and +1 under all retained ones, satisfying the active-irrep condition (Supplementary Table 3). Its linear basis function is $P_z$, a polar displacement along z confirming its identity as a zone-centre polar mode. In $(PA)_2FAPb_2I_7$, however, $B_{1u}$ is not the primary order parameter. The dominant lattice instability is the zone-boundary octahedral rotational and tilt mode. The polar $B_{1u}$ displacement is induced secondarily through the trilinear coupling term $X_2^+ \otimes X_3^- \otimes \Gamma_4^-$ in the Landau free energy. This is the defining signature of hybrid improper ferroelectricity[1,2].

The lattice dynamical properties of the high-symmetry *Pnma* phase were analysed using DFT to identify the distortion mode pathway leading to the polar $Pna2_1$ phase. Phonon dispersion calculations (Main Figure 2c) reveal the presence of imaginary frequencies at both the Brillouin-zone center (Γ) and zone-boundary points (S and X), particularly the X point (½, 0, 0). At the Γ point, irreducible representation (irrep) analysis identifies several unstable polar modes ($B_{1u}$, $B_{2u}$, and $B_{3u}$) that are, in principle, capable of generating macroscopic polarization. However, these instabilities are relatively weak compared to the dominant zone-boundary phonon modes. In addition, the prominent unstable Γ-point mode belongs to non-polar (gerade-type) irreps $B_{2g}$ and $B_{3g}$, while the polar instabilities are smaller in magnitude at Γ-point, indicating that the primary lattice instability at Γ does not directly break inversion symmetry. Therefore, the Γ-point instabilities alone cannot account for the symmetry reduction from Pnma to $Pna2_1$. In contrast, the phonon spectrum at X-point displays strongly unstable modes with large imaginary frequencies and characteristic degeneracies of zone-boundary distortions. The substantial instability at X indicates that the primary structural distortion is driven by zone-boundary phonon as well rather than solely by a Γ-point polar mode. The symmetry mode decomposition of the distorted structure at X-point, we find that the zone-boundary distortion is led by rotational mode $X_2^+$ and tilt mode $X_3^-$ (Supplementary Table 2, Main Figure 2d)

Within a Landau framework, with coupled order parameters, the $X_2^+$ and $X_3^-$ distortion acts as the primary order parameter, while the $\Gamma_4^-$ point polar mode is secondary and coupled to it. This hierarchy is characteristic of improper ferroelectricity, in which a dominant nonpolar lattice instability ($X_2^+$) induces polarization through coupling to a secondary polar ($\Gamma_4^-$) mode. Here, the $X_2^+$ distortion indirectly breaks inversion symmetry and stabilizes the polar $Pna2_1$ phase

In the centrosymmetric *Pnma* structure, the $FA^+$ cations and one iodine atom occupy the 4c Wyckoff site with $y = 1/4$ or $3/4$, constrained by a mirror plane $m_z$(i.e., $\sigma_{xy}$) perpendicular to the *b* axis (Main Figure 2a-b). This symmetry fixes the molecule on the mirror plane, limiting its orientational freedom within the $PbI_6$ octahedral framework. However, FA cations are dynamically disordered at ambient temperature, and even slight reorientations can perturb the surrounding lattice through hydrogen bonding and steric effects. Such deviations can couple to cooperative octahedral rotations corresponding to a zone-boundary mode misalignment/distortion (Main Figure 2d). Once this rotational distortion develops, the mirror symmetry is lost, allowing the FA cation to shift off the plane and break inversion symmetry, thereby lowering the symmetry from *Pnma* to $Pna2_1$. This coupling between molecular dynamics and octahedral rotations provides a natural pathway for the symmetry reduction and thereby invoking a strong hybrid improper ferroelectricity.

**Table 2:** Octahedral distortions at zone boundary X-point accounting for out-of-plane tilt ($X_3^-$) and in-plane rotation ($X_2^+$)

| Phonon branch at X-point | Out-of-plane tilt (Octahedron-1, Octahedron-2) | In-plane rotation (Octahedron-1, Octahedron-2) | Octahedral Distortion |
|---|---|---|---|
| 1 or 2 | 175°, 175° | 152°, 147° | $X_2^+$ (Pure rotation) |
| 3 or 4 | 175°, 160° | 149°, 149° | $X_3^-$ & $X_2^+$ (Tilt & rotation) |
| 5 or 6 | 175°, 175° | 152°, 149° | $X_2^+$ (Pure rotation) |
| 7 or 8 | 172°, 166° | 147°, 147° | $X_3^-$ & $X_2^+$ (Tilt & rotation) |

**Table 3.** Character Table of $D_{2h}$ and $C_{2v}$ point groups, corresponding to space groups Pnma (62) to $Pna2_1$ (33), respectively

**D2h Point Group (Pnma)**

| | E | $C_2(z)$ | $C_2(y)$ | $C_2(x)$ | i | σ(xy) | σ(xz) | σ(yz) | linear/rotations | quadratic |
|---|---|---|---|---|---|---|---|---|---|---|
| $A_g$ | 1 | 1 | 1 | 1 | 1 | 1 | 1 | 1 | | $x^2$, $y^2$, $z^2$ |
| $B_{1g}$ | 1 | 1 | -1 | -1 | 1 | 1 | -1 | -1 | Rz | xy |
| $B_{2g}$ | 1 | -1 | 1 | -1 | 1 | -1 | 1 | -1 | Ry | xz |
| $B_{3g}$ | 1 | -1 | -1 | 1 | 1 | -1 | -1 | 1 | Rx | yz |
| $A_u$ | 1 | 1 | 1 | 1 | -1 | -1 | -1 | -1 | | |
| $B_{1u}$ | 1 | 1 | -1 | -1 | -1 | -1 | 1 | 1 | z | |
| $B_{2u}$ | 1 | -1 | 1 | -1 | -1 | 1 | -1 | 1 | y | |
| $B_{3u}$ | 1 | -1 | -1 | 1 | -1 | 1 | 1 | -1 | x | |

**$C_{2v}$ Point Group ($Pna2_1$)**

| | E | $C_2(z)$ | $\sigma_v(xz)$ | $\sigma_v(yz)$ | linear/rotations | quadratic |
|---|---|---|---|---|---|---|
| $A_1$ | 1 | 1 | 1 | 1 | z | $x^2$, $y^2$, $z^2$ |
| $A_2$ | 1 | 1 | -1 | -1 | Rz | xy |
| $B_1$ | 1 | -1 | 1 | -1 | x, Ry | xz |
| $B_2$ | 1 | -1 | -1 | 1 | y, Rx | yz |

## Moiré superlattices

## Supplementary Note 6

Subsequently, evidence for rotational disorder is further supported by Laue diffraction pattern. Precession images of $(PA)_2FAPb_2I_7$ simulated from the SCXRD along (hk0) clearly shows discrete diffraction spots, indicating inter-layer alignment along the planes (Main Figure 2e). However, along the (h0l) and (0kl) planes, streaking or elongation of diffraction spots signifies disorder or misalignment between layers, possibly along the layer stacking direction (Main Figure 2f-g). This anisotropic diffuse scattering signifies reduced coherence and local rotational misalignment along the (010) direction, which coincides with the van der Waals stacking axis of the layered structure. Such behavior is consistent with local $X_2^+$ and $X_3^-$ type rotational distortions between adjacent layers, which preserve in-plane order to some extent but disrupt long-range stacking coherence. These patterns are consistent with strain or stacking defects that occur during crystal growth, as observed in 2D perovskites[3]. Optical and electron microscopy further reveal the presence of twin-like features within individual crystals. Visible light microscopy imaging reveals the presence of twin-like features within individual crystals, consistent with ferroelastic domain formation (Main Figure 2h). These domains are crystallographically related and share nearly identical lattice parameters but differ in orientation through a symmetry operation that is present in the higher-symmetry parent phase and lost upon distortion. The pseudo-merohedral twinning observed in single-crystal diffraction therefore reflects intrinsic symmetry breaking, producing multiple energetically equivalent orientation variants. Such behavior is characteristic of collective lattice distortions, particularly rotational instabilities, rather than a simple uniform polar shift, and is naturally expected when a zone-boundary rotational mode condenses and generates degenerate structural states. SEM imaging of crystal and of exfoliated single crystals confirms that the crystals retain sharp facets and well-defined morphology, ruling out surface degradation or mechanical damage as the origin of the observed twinning (Main Figure 2i).

Direct real-space and reciprocal-space evidence for rotational misalignment is obtained from transmission electron microscopy. HRTEM confirms the presence of such twinning at the nanoscale, complementing the optical microscopy observations and revealing coherent domain boundaries consistent with ferroelastic switching in the distorted perovskite lattice (Main Figure 2j). Selected area electron diffraction (SAED) patterns exhibit secondary diffraction spots that are offset relative to the primary lattice (Main Figure 2k). Such an offset is indicative of weak rotational misalignment between adjacent domains or layers[4]. The relative twist angle between the splitting diffraction spots was determined to be ~5.17°. The corresponding moiré periodicity calculated using $\mathrm{L} = \mathrm{a}/(2\sin(\theta/2))$is ~2.55 nm. This rotational mismatch gives rise to a moiré superlattice, which is directly observed in the HRTEM image, indicating interference pattern arising from the overlay of two slightly misaligned periodic lattices (Main Figure 2l). The periodicity of this modulation was measured to be ~2.5 nm, in excellent agreement with the value predicted from the twist angle above. The zoomed-in view of the HRTEM image (Main Figure 2m) exhibits a hexagonal pattern, consistent with the formation of a moiré superlattice due to twisted stacking[5,6,7]. The corresponding inverse fast Fourier transform (IFFT) exhibits noticeable spatial modulation and curvature of the fringes (Main Figure 2n). The observed wavy moiré lattice fringes can be attributed to the combined effect of strain and twist-angle variation associated with twin domain boundaries. The presence of orthorhombic twin domains provides a natural origin for such effects. Interfaces between twin domains introduce lattice mismatch and shear distortions, leading to localized strain and slight rotational misalignment between adjacent regions.

2D perovskites serve as a versatile material system for constructing such twisted layers. However, the existing synthesis methods have not succeeded in producing 2D perovskite crystals with precisely controlled twist angles, which are essential for creating superlattices. Nevertheless, artificial twisted structures of 2D perovskite have been constructed using 2D RP perovskites as templates[34]. Similarly, Ab initio simulation elucidates the mechanism behind the twisted layers in 2D Ruddlesden-Popper perovskites and predicts the emergence of moiré excitons[8].

## Piezoelectric and Ferroelectric Properties

### Supplementary Note 7

Although PFM is the most prevalent method for characterizing piezoelectric materials, it is susceptible to electrostatic, mechanical or chemical artifacts[9,10].To eliminate these unwanted effects, a combination of electrostatic blind spot method, substrate subtraction method and off-resonance imaging have been used[11,12,13,14]. Consequently, minimal electrostatic contribution or background noise could be detected from the underlying Si substrate (Supplementary Figure 8). Furthermore, electrostatic interference can also be precluded by using large reversal voltages to realize phase switching. As shown earlier in Main Figure 3g-h and Supplementary Figure 7, such measurements also help to substantiate intrinsic piezoelectricity in 2D $(PA)_2FAPb_2I_7$. It has been shown that a longer and stiffer cantilever can further diminish the electrostatics in PFM measurements. This is because a larger tip length is known to attenuate the capacitive coupling between the tip and the surface and electrostatic contributions are inversely proportional to the stiffness of the cantilever[12,15]. The above results were therefore validated using a stiffer probe with longer cantilever length (see Methods). As shown in Supplementary Figure 9, similar piezo and phase contrast results are obtained from the 2D $(PA)_2FAPb_2I_7$ samples using this alternative cantilever, suggesting minimal electrostatic contribution.

### Supplementary Note 8

Polarization of the material directly impacts the strength of piezoelectricity and hence the piezoelectric coefficient. They are related through $d_{33} \propto \varepsilon P_r$ where $\varepsilon$ is the dielectric permittivity and $P_r$ is the remnant polarization.[60] It is well known that the A-site cation and its rotation degree of freedom, in perovskite structure is the main origin for polarization and hence piezo response[16]. Another crucial parameter is the breaking of the inversion symmetry of the crystal and it usually occurs when the B-site cation moves away from the center of octahedron, creating dipole along the process and contributing to piezoelectric effect in halide perovskite[17,18,19]. This is largely dependent on the size and lengths of A-size cation which lies within the octahedral void. In 2D $(PA)_2FAPb_2I_7$, it is the orientation of $FA^+$ organic spacer cations that can either make the structure centrosymmetric or non-centrosymmetric (Main Figure 2a-b).

### Supplementary Note 9

The rotational misalignment of layers during single crystal growth allows local breaking of center of inversion of the crystal lattice and subsequently induces polar deformations in the $PbI_3^{-2}$ cage, ultimately resulting in elevated piezoelectric amplitude and coefficient in nominally centrosymmetric layered systems. This observation aligns with a recent work on twisted bilayer hexagonal boron nitride (h-BN), which reports that pattern formation can induce periodic in-plane strain and charge redistribution through the intrinsic piezoelectric effect, thereby establishing a direct link between interlayer twisting and enhanced piezoelectric response[20]. Furthermore, long chains of PA organic cations culminate in augmented distance between the $PbI_6^{-4}$ octahedral sheets allowing for more space to contract or expand under an applied external field, which likely also contributes to the high piezoelectricity in 2D $(PA)_2FAPb_2I_7$. The density and configuration of polar domains (regions with local symmetry breaking) can change with thickness. In some materials, thinner films may have fewer or smaller non-centrosymmetric domain regions, or the domains may become less stable, further reducing the piezoresponse[21,22]. The reduction of the PFM signal with decreasing thickness reflects both the diminishing contribution of locally non-centrosymmetric regions and the dominance of the average centrosymmetric structure as the system approaches the atomic limit. This trend is a direct manifestation of the interplay between local symmetry breaking and the overall structural symmetry of the material. However, it should be noted that in atomically thin layered films, the depolarization field becomes more significant, which can also suppress the polar order responsible for piezoelectricity, thereby reducing the measurable PFM signals[23].

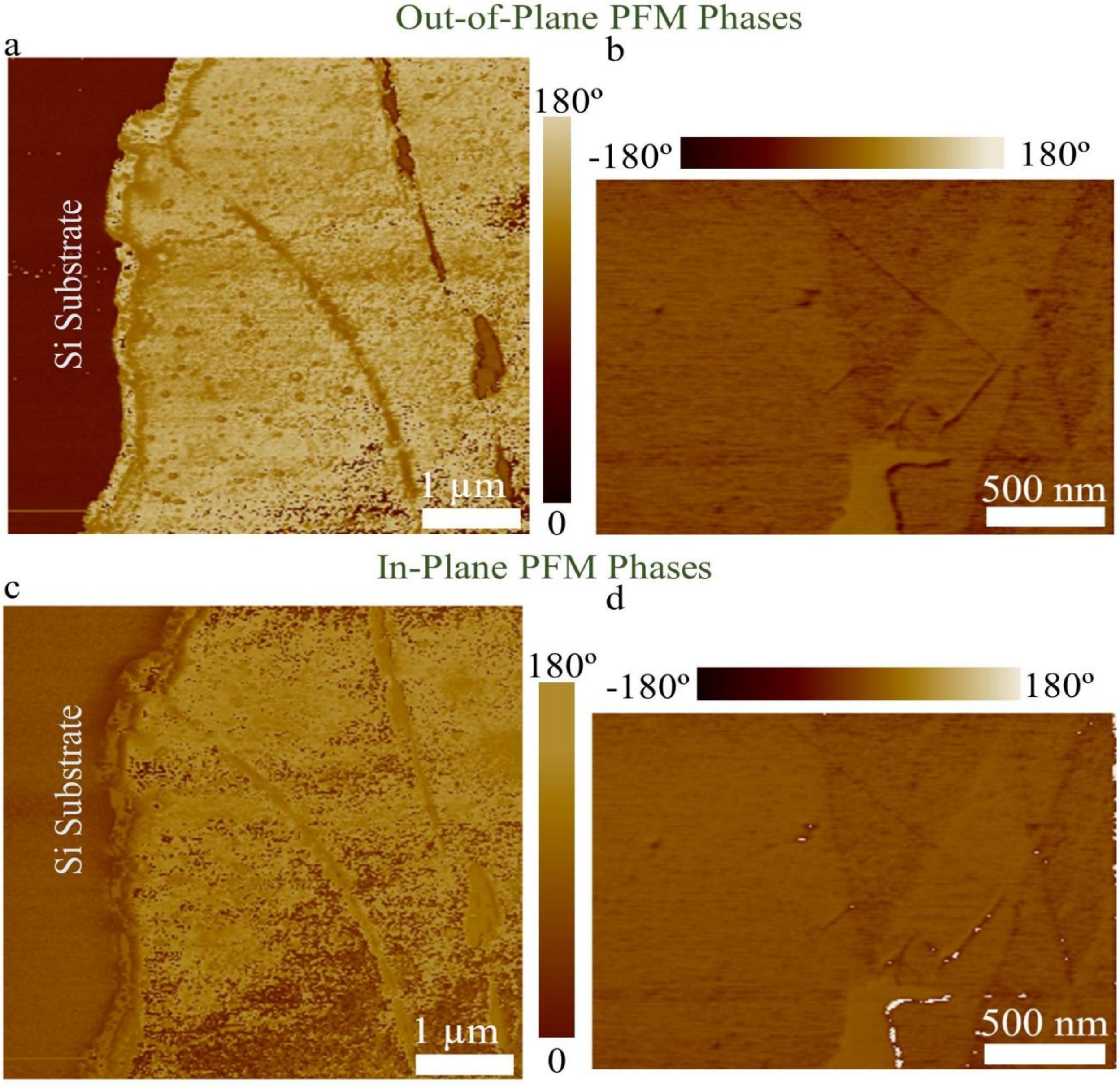

**Figure 4: (a-b)** Out-of-plane and **(c-d)** in-plane PFM Phase images of bulk and atomically thin layers of the samples shown in main text. Clear Phase contrast is obtained compared to the Si substrate for both the bulk and thin layers in both vertical and horizontal direction.

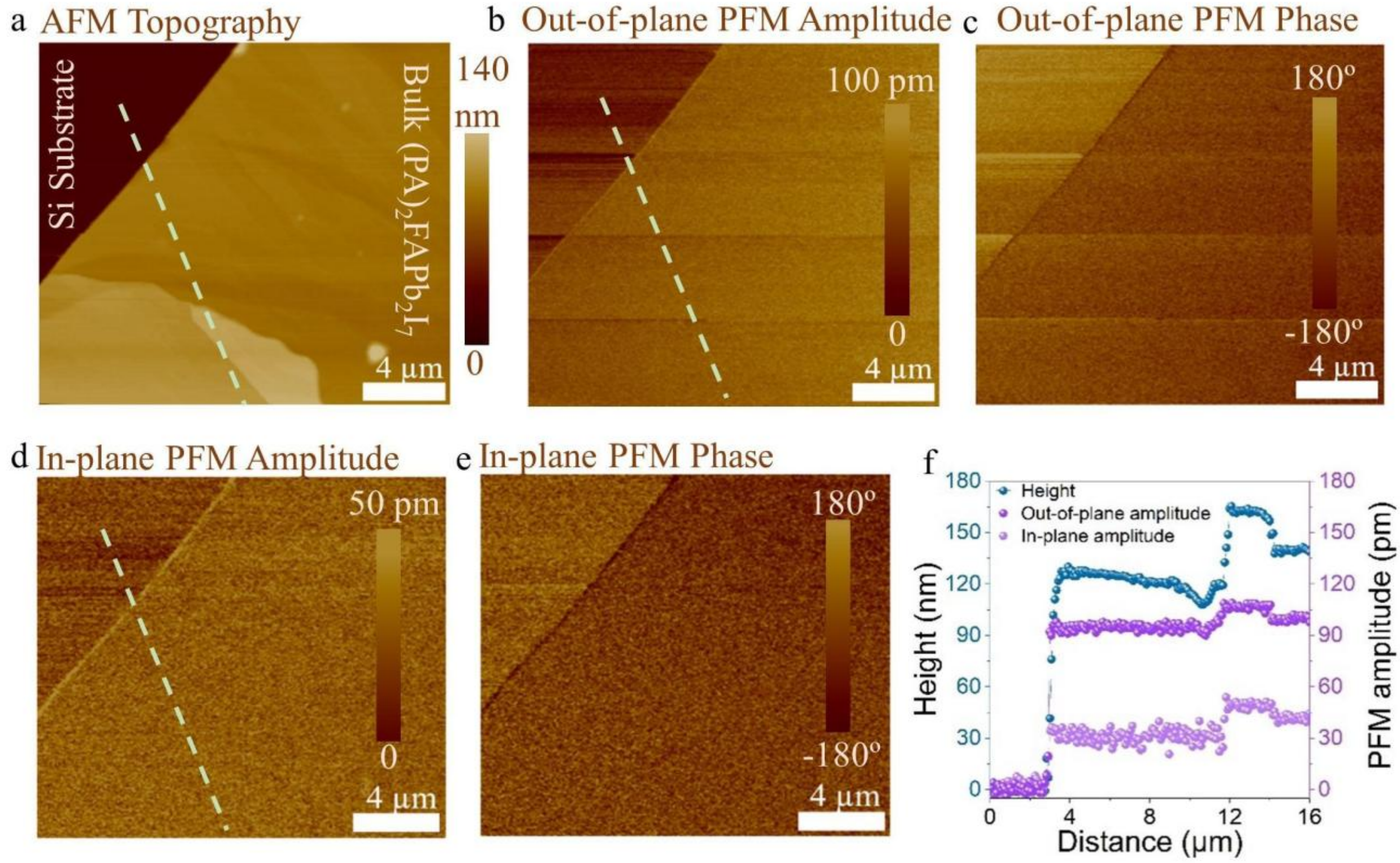

**Figure 5:** PFM of another 2D $(PA)_2FAPb_2I_7$ bulk perovskite to ensure repeatability. **(a)** AFM topography of bulk and multilayer perovskite on Si substrate. Thickness ranges between 100nm to 165nm. **(b-c)** Corresponding out-of-plane and **(d-e)** in-plane PFM amplitude and phase, revealing RT piezoelectricity. **f** Height and amplitude line profile along the green dotted line shown in **(a)**, **(b)** and **(d)**. Piezo response varies with the sample thickness.

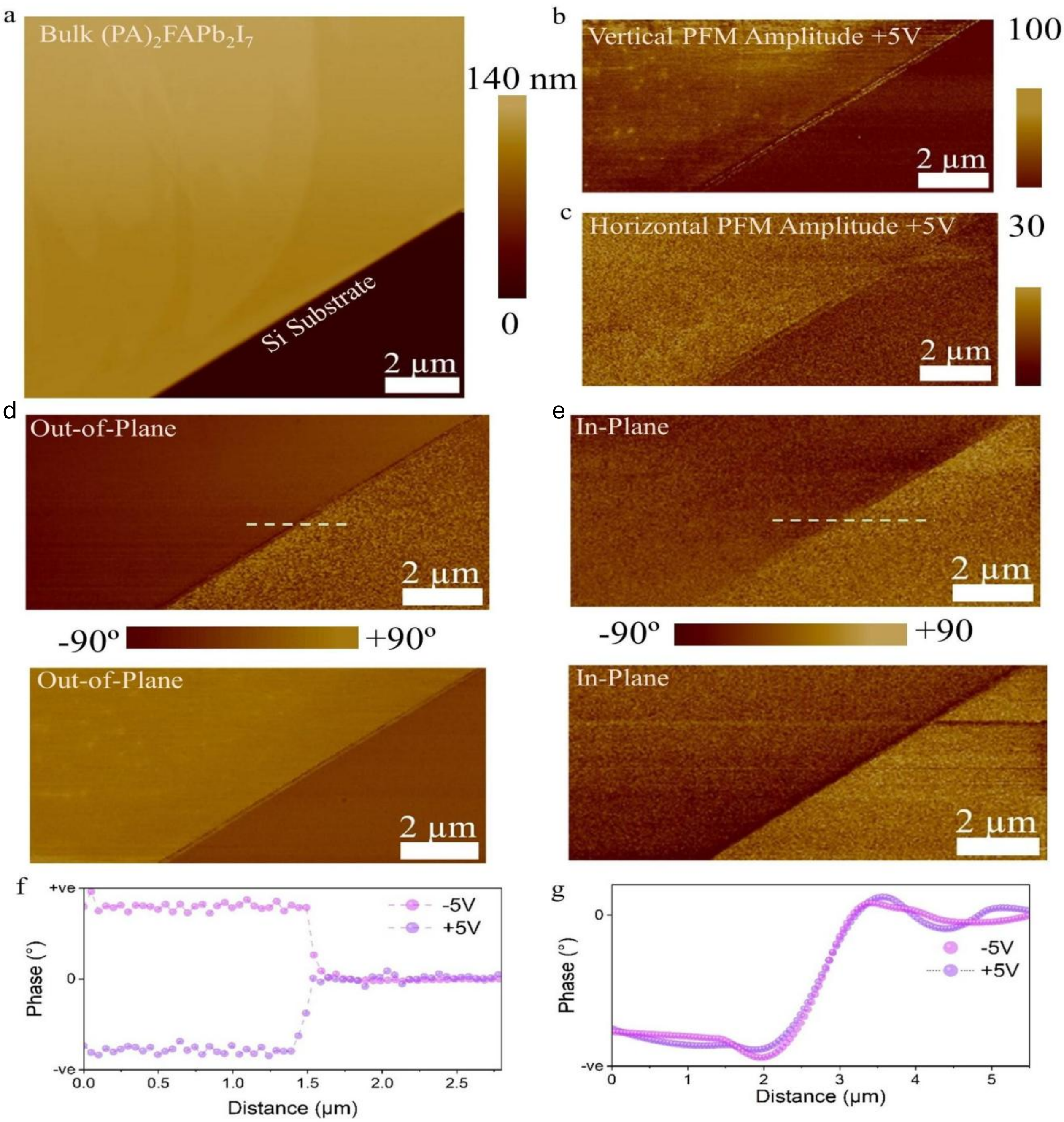

**Figure 6:** Evidence of ferroelectricity in 2D $(PA)_2FAPb_2I_7$: **(a)** AFM Topography of a exfoliated bulk perovskite sample with thickness of around 100nm, **(b-c)** Vertical and horizontal piezo amplitude of the selected region in a, obtained from piezo-response force microscopy (PFM) measurements. Similar piezo response is obtained to that of the sample in main text. **(d-e)** Corresponding PFM phase images and **(f, g)** corresponding profiles along the green dotted line. The phase signal is recorded with both a voltage bias of +5 V and -5 V to highlight that a switch in the orientation of the polar domains of the material in the out-of-plane direction is induced by changing the sign of the voltage bias applied to the tip which proves ferroelectricity. However, this has not been observed along the in-plane direction.

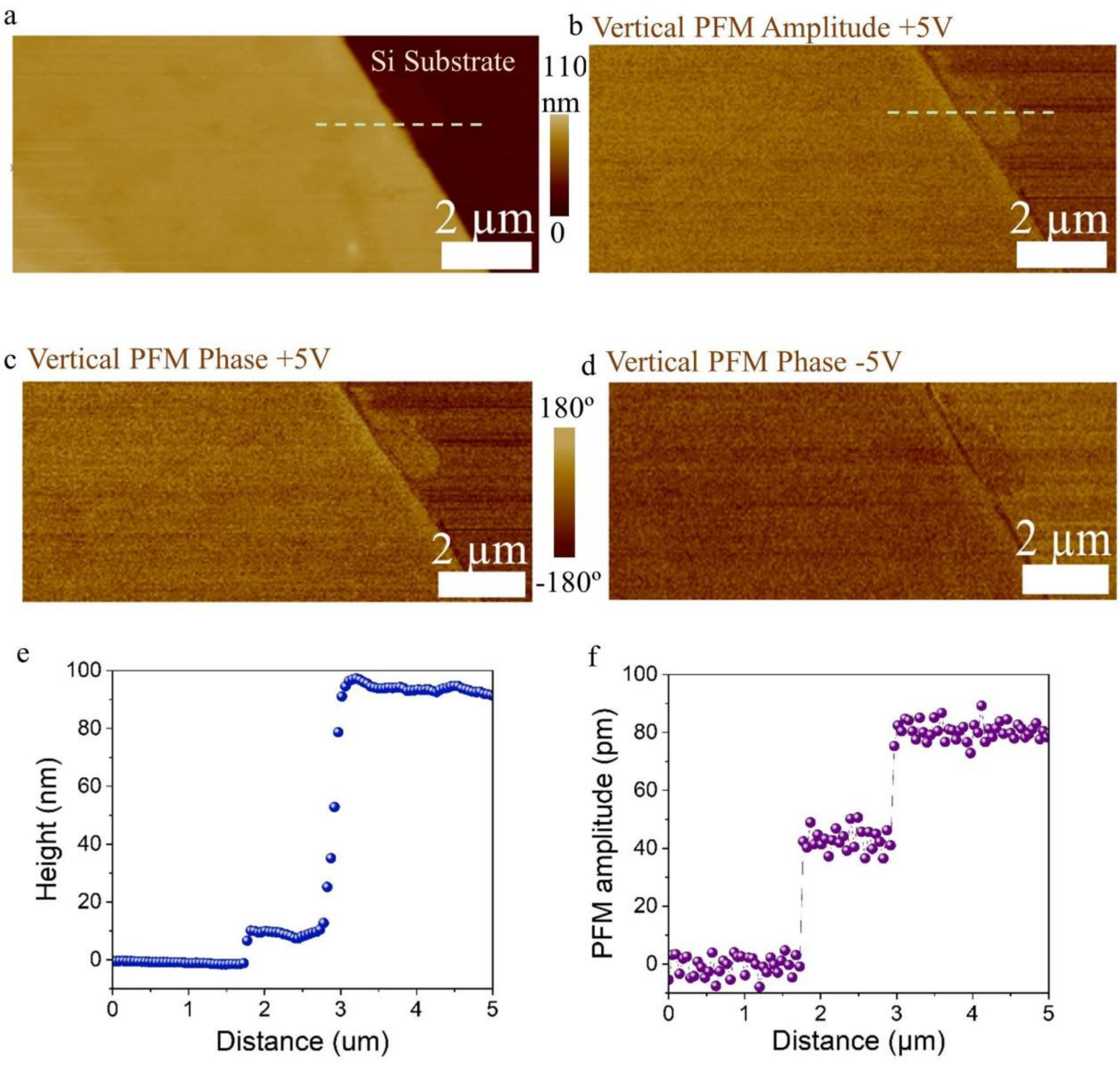

**Figure 7:** Ferroelectric switching in another 2D perovskite sample to ensure repeatability. **(a)** AFM topography of 2D $(PA)_2FAPb_2I_7$ consisting of bulk and atomically thin layers. **(b)** Corresponding out-of-plane PFM amplitude. **(c-d)** Out-of-Plane PFM phase at +5 and -5V respectively. **(e-f)** Line Profile along the dotted line in **(a)** and **(b)** showing the AFM and PFM amplitude profile. Thin layer obtained is 7L.

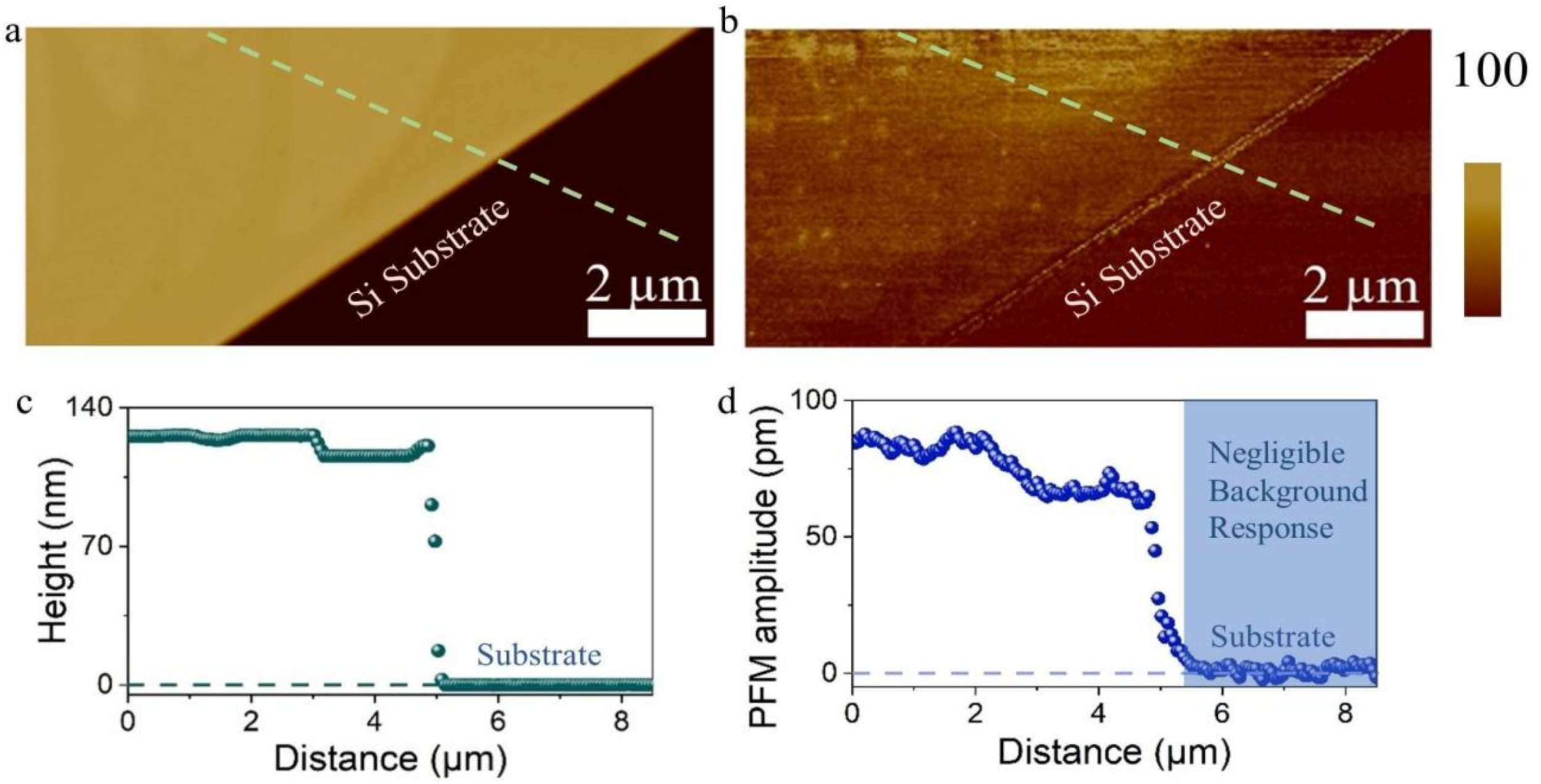

**Figure 8: (a-b)** AFM topography and corresponding PFM amplitude of a bulk (ca. 130nm) 2D $(PA)_2FAPb_2I_7$ perovskite at +5V. **(c-d)** AFM and PFM Line profile along the green dotted line in a and b. Negligible background response can be recorded from the substrate.

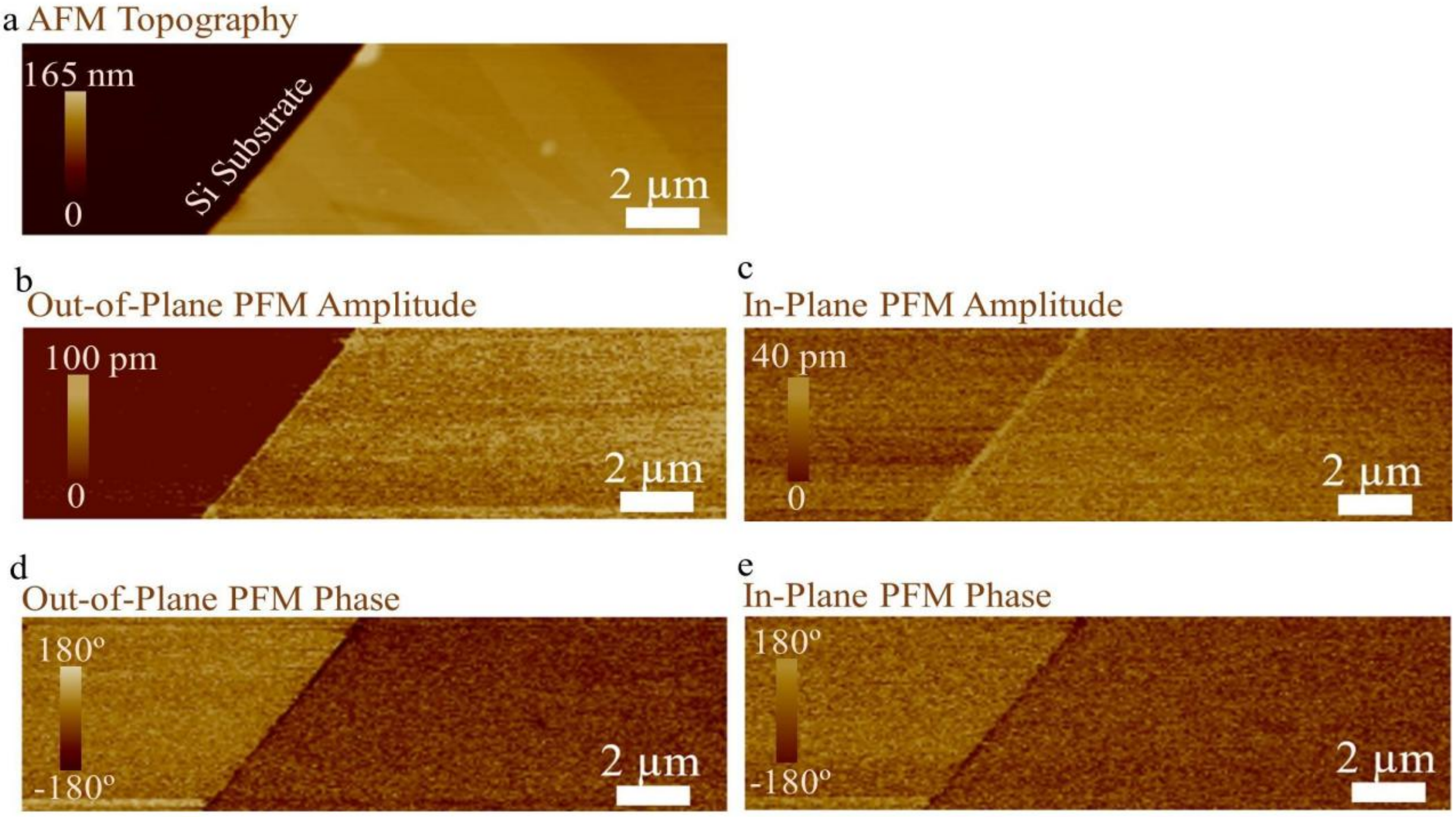

**Figure 9:** PFM of 2D perovskite $(PA)_2FAPb_2I_7$ using a stiffer cantilever. **(a)** Topography of a bulk perovskite. **(b-c)** corresponding out-of-plane and **(d-e)** in-plane PFM amplitude and phase captured using a stiff cantilever (DDESP-FM-V2) with K=6 N/m. Similar PFM amplitude and phase variations can be obtained compared to the previously used cantilever. This helps to rule out electrostatics.

**Table 4:** Summary of Piezoelectric Coefficients for various materials

| Piezoelectric Materials | | $d_{33}$ (pm/V) | $d_{31}$ (pm/V) | Comments | Reference |
|---|---|---|---|---|---|
| Van der Waals Materials | α-$In_2Se_3$ | 0.34 | | Monolayer, PFM Measurements | 10 |
| | 3R-$MoS_2$ | 0.27 | 0.21 | Bulk/Multilayer, DFT Calculations | 24 |
| | 3R-$WS_2$ | 0.30 | 0.14 | | |
| | 3R-$MoSe_2$ | 0.45 | 0.1 | | |
| | 3R-$WSe_2$ | 0.35 | | | |
| | MoSSe | 5.248 | 0.109 | Bilayer, DFT Calculations | 25 |
| | WSSe | 5.319 | 0.017 | Bilayer, DFT Calculations | |
| | $SnS_2$ | 5 | | Thin Layer, PFM Measurements | 26 |
| | Graphene | | 0.3 | | |
| | VSSe | 4.92 | | Multilayer, First Principle | 27 |
| | g-$C_3N_4$ | 1 | | Nanosheets, PFM Measurements | 28 |
| | **This work $(PA)_2FAPb_2I_7$** | **19.9** | **3.54** | **Bulk and Thin Layers, PFM** | |
| III-V Semiconductors | Wurtzite GaN | 2.8-3.7 | | Bulk and/or Thin Films, Experiential Deductions | 29 |
| | Wurtzite AlN | 5.1 | | | |
| | InP | 0.5 | | | 30 |
| | GaP | 0.82 | | | |
| | GaAs | 1.54 | | | |
| | InAs | 0.72 | | | |
| Janus Chalcogenides | ZrSeS | 0.34 | | Bulk, DFT Calculations | 28,31 |
| | HfSeS | 0.21 | | | |
| | MoSO | -5.23 | | | |
| | MoSeS | 2.58 | | | |
| | WSeO | -6.53 | | | |
| | WSO | -2.86 | | | |
| | WSeS | 2.56 | | | |
| | MoSTe | 5.7 | | Multilayer, First Principle | |
| | $Ga_2SSe$ | 0.3 | 0.07 | Thin layers, DFT Calculations | |
| | $Ga_2STe$ | | 0.25 | | |
| | Ga2SeTe | | 0.21 | | |
| | $In_2SSe$ | | 0.18 | | |
| | $In_2STe$ | | 0.25 | | |
| | $In_2SeTe$ | | 0.13 | | |
| | $GaInTe_2$ | | 0.32 | | |
| | $GaInS_2$ | | 0.38 | | |
| | $GaInSe_2$ | | 0.46 | | |
| 2D and 3D Perovskite | 2D Homochiral PbI4 | 3 | | Bulk and Thin Films, PFM and First Principles | |
| | MDABCO-$NH_4I_3$ | 12.8 | | | |
| | 2D $MAPbI_3$ | 10.8 | | | |

| | $MAPbI_3$ | 25 | | | 32, 33 |
|---|---|---|---|---|---|
| | $MASnI_3$ | 20.8 | | | |
| | $MAPbBr_3$ | 16 | | | |
| | $MAPbCl_3$ | 3.4 | | | |
| | $FAPbBr_3$ | 25 | | | |
| | $CsPbI_3$ | 15 | | | |
| | $CsPbBr_3$ | 9.3 | | | |

## Excitonic Emission Dynamics

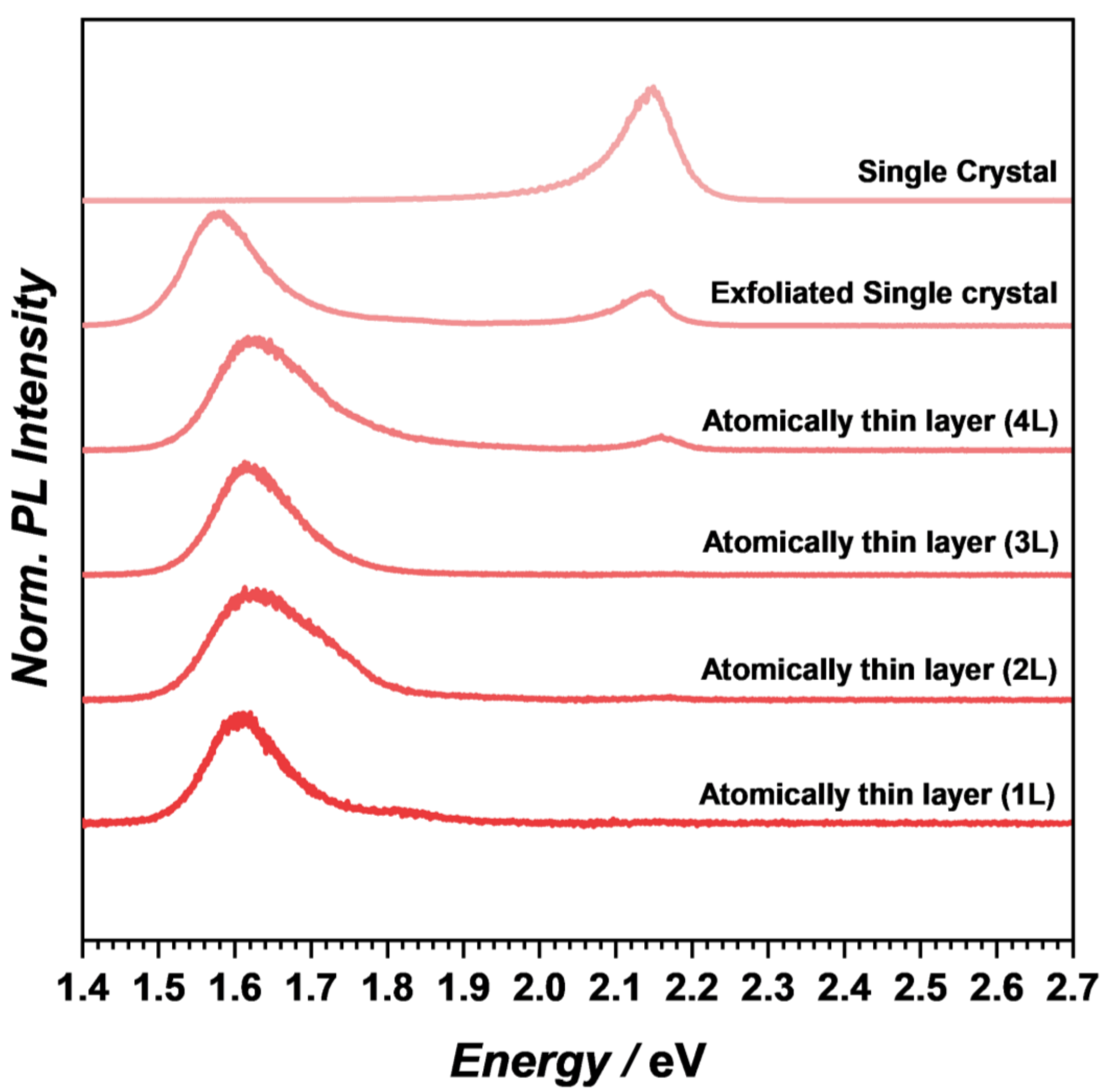

**Figure 10:** Photoluminescence spectra of single crystal, exfoliated single crystal bulk and atomically thin layers of $(PA)_2FAPb_2I_7$ perovskite at room temperature.

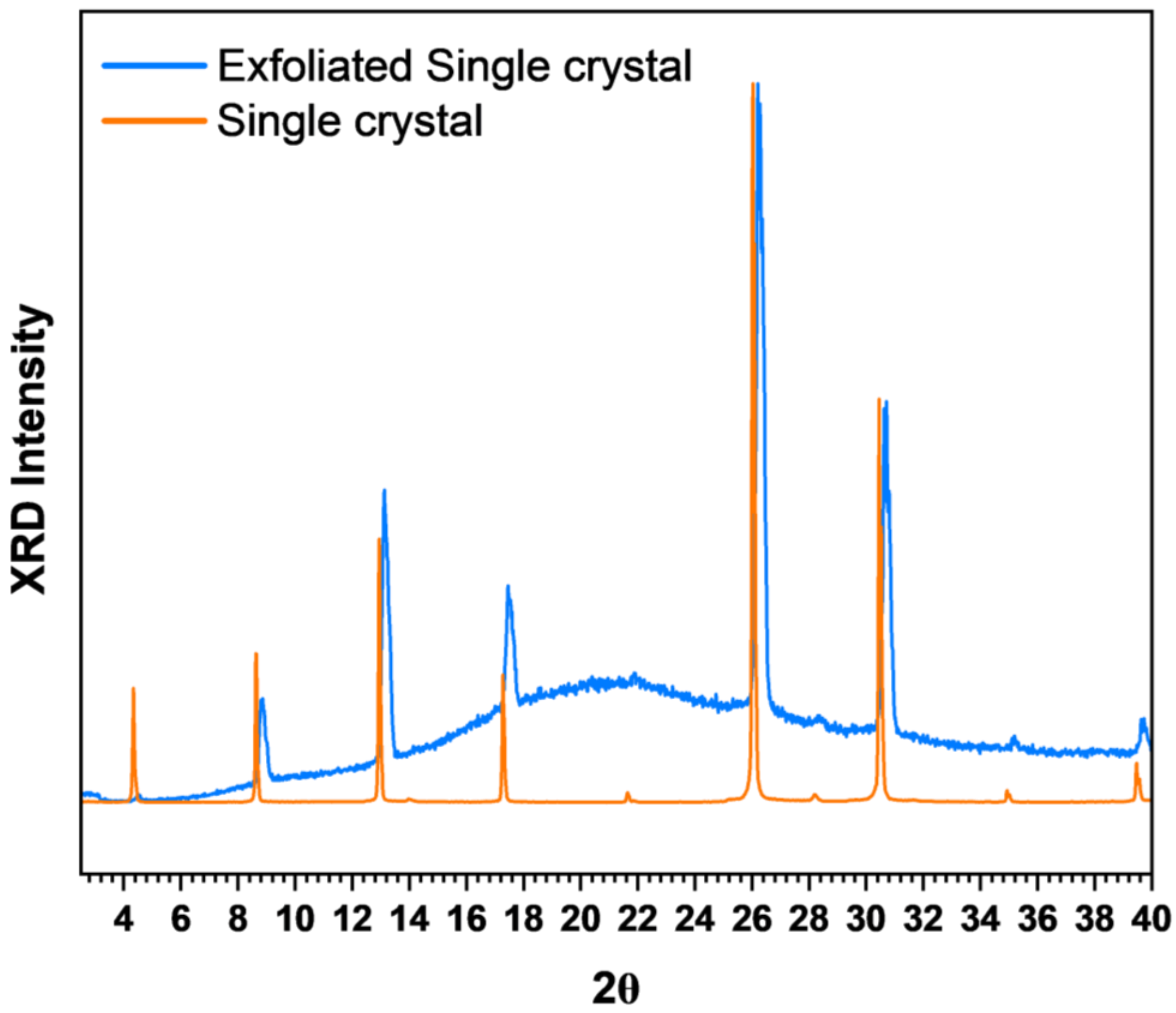

**Figure 11:** X-Ray diffraction profile of single crystal and exfoliated single crystal, indicating contraction of lattice upon exfoliation of quasi-2D piezoelectric $(PA)_2FAPb_2I_7$ single crystal flakes.

## Supplementary Note 10

The steady-state PL properties of the as-obtained exfoliated single crystals and atomically thin layers of $(PA)_2FAPb_2I_7$ are further studied to investigate the anomalous emissive dynamics of $(PA)_2FAPb_2I_7$ (Supplementary Figure 10). The PL emission in the single crystals is predominantly from the band-to-band excitonic transition, corresponding to the band gap of the system *ca.* 2.15eV (high-energy Peak). Along with the main excitonic PL emission i.e. high-energy Peak, a stoke-shifted broad PL sideband (*ca.* 1.65 eV) low-energy Peak emerges from an exfoliated single crystal sample. As we further move to atomically thinner layers of $(PA)_2FAPb_2I_7$, the high-energy Peak emission starts to diminish and completely disappears with only a single PL emission from low-energy Peak, noticeable for atomically thin layer 2L. As evidenced by layer-dependent PL analysis, we hypothesize that the commonly referred edge state emission, might be emerging from the thinner layers around the periphery of the crystal edge, which aligns with the proposition that edge state occurrence in 2D RP perovskite is sensitive to the thickness of single crystal[34].

The intrinsic mechanism that triggers the low-energy $X_5$ is subject to further investigation. The XRD-profile (Supplementary Figure 11) of exfoliated single crystal of $(PA)_2FAPb_2I_7$ negates the presence of self-form 3D perovskite $FAPbI_3$. However, there is a slight but uniform shift in all the XRD peaks, indicating a uniform contraction/relaxation of the lattice structure upon exfoliation. To retaliate, as per the theoretical hypothesis by Hong et al[35], there is a possibility that upon application of stress, the internal electric field induced in the piezoelectric domain regions of $(PA)_2FAPb_2I_7$ can propel a stress-induced structural deformation on the surface due to ferroelastic properties of non-centrosymmetric structures. Whether or not, the piezoelectricity-induced structural deformation leads to low-energy $X_5$ emission, and the impetus mechanism behind it is further extensively discussed in subsequent sections.

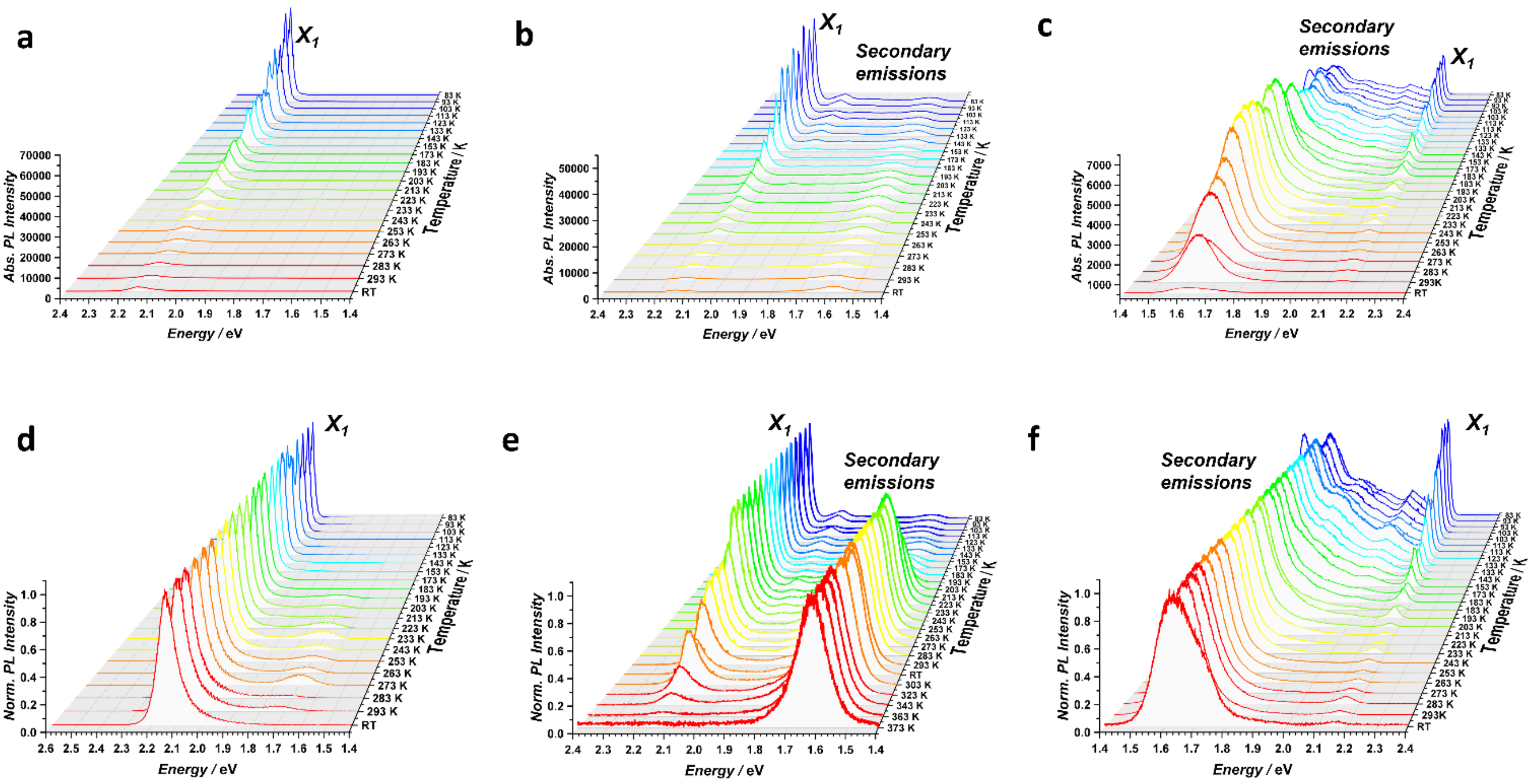

**Figure 12:** Temperature-dependent PL spectra of **(a)** single crystal, **(b)** exfoliated single crystal, and **(c)** atomically thin layer of $(PA)_2FAPb_2I_7$ perovskite, showing the evolution of distinct absolute PL intensities from high temperature (373 K) to cryogenic temperature (83 K). **(d,e,f)** are corresponding normalized PL intensities.

To investigate the temperature-dependent emissive dynamics of our single crystals, we performed PL measurements from 373 K to 83 K. For the as-synthesized single crystals (Supplementary Figure 12a), temperature-dependent PL spectra displayed a single high-energy peak ($X_1$) at room temperature, with additional low-energy peaks (secondary emissions) emerging below 293 K and diminishing below 193 K (Supplementary Figure 12d). In exfoliated bulk crystals as observed in Supplementary Figure 12b, both $X_1$ and secondary emission peak ((*ca.* 1.65 eV)) are present at room temperature; on heating to 373 K, only a broad secondary emission remains, consistent with a structural phase transition near *ca.* 349 K (Supplementary Figure 12e). As temperature decreased below this transition, $X_1$ reappeared and sharpened, analogous to a zero-phonon line; while secondary emission peak's intensity reduced below 173–203 K. Notably, a distinct intermediate secondary emission peak $X_2$ is observed and secondary emission peak resolves into multiple equidistant PL peaks $X_3$, $X_4$, and $X_5$ 243 K (Main Figure 4a). Similar behavior is seen in atomically thin 4L samples (Supplementary Figure 12c), where the broad low-energy emission splits into multiple evenly spaced peaks, upon cooling to 83 K. As evidenced by layer-dependent PL analysis, we hypothesize that the commonly referred edge state emission, might be emerging from the thinner layers around the periphery of the crystal edge, which aligns with the proposition that edge state occurrence in 2D RP perovskite is sensitive to the thickness of single crystal[34]. Temperature-dependent PL spectra acquired from another exfoliated single crystal sample (Supplementary Figure 13), illustrating the evolution of spectral features from room temperature to cryogenic conditions. Multiple peaks emerge at low temperature, and the results are reproducible across samples.

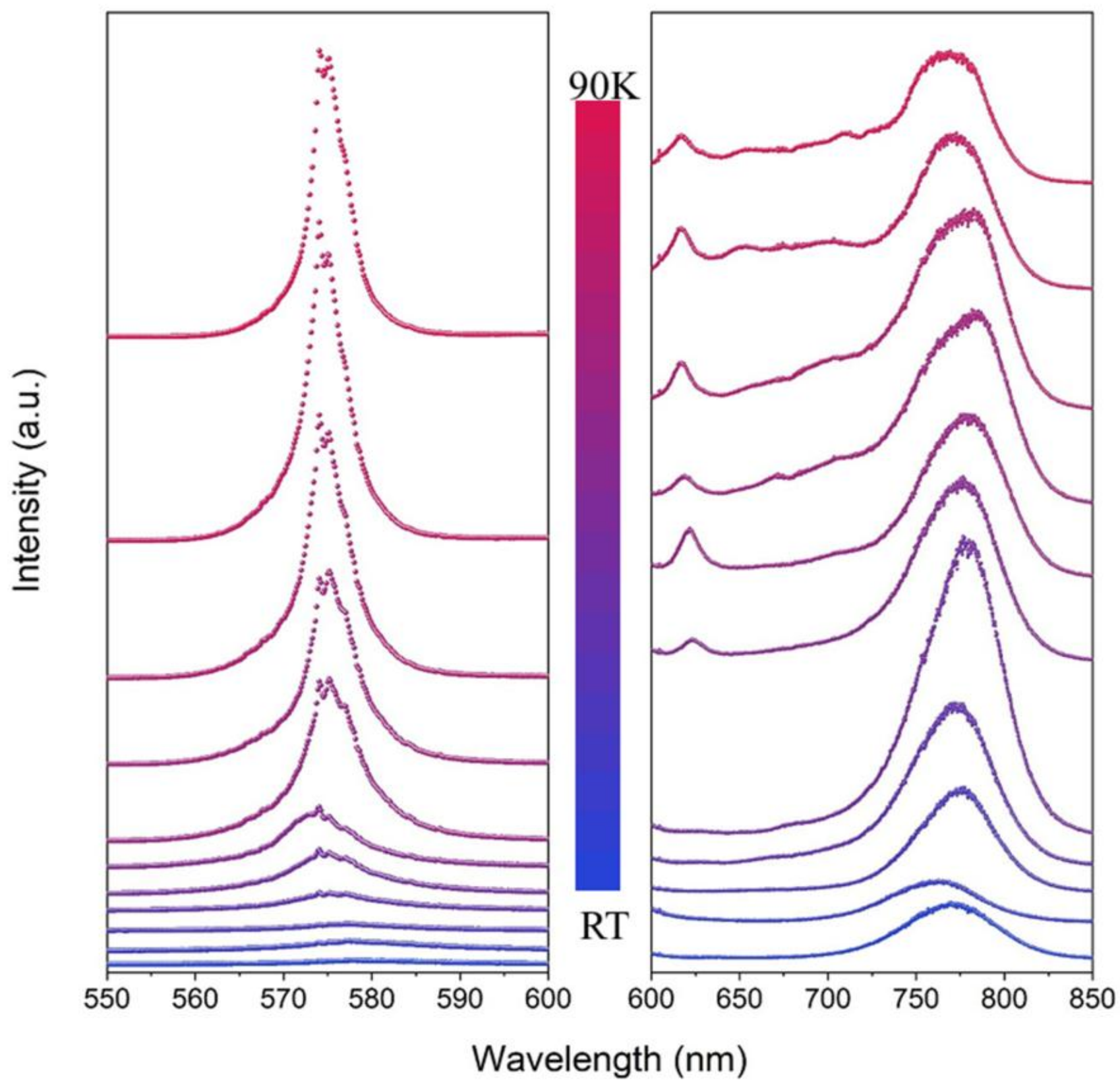

**Figure 13:** Temperature-dependent PL spectrum from another exfoliated single crystal sample, showing the evolution of spectral peaks from RT to low temperature. Repeatable results can be obtained with emergence of multiple peaks at low temperature.

## Supplementary Note 11

Acoustic phonon, longitudinal optical phonon, and impurity scattering mechanisms each impart characteristic temperature dependencies to the PL linewidth. Using the Bose-Einstein statistics, the resultant temperature-dependent PL linewidth can thus be formulated as the sum of these contributions, as follows[36] .

$$\Gamma = \Gamma_0 + \Gamma_{ac} + \frac{\Gamma_{LO}}{e^{\frac{E_{LO}}{kT}} - 1} + \Gamma_{imp} \quad (1)$$

The first term represents inhomogeneous thermal broadening $\Gamma_0$, followed by contributions from acoustic phonon $\Gamma_{ac}$ and longitudinal optical (LO) phonon scattering $\Gamma_{LO}$, with the last term describing electron-phonon interactions from ionized impurities $\Gamma_{imp}$. The LO phonon term accounts for the Fröhlich interaction arising from Coulomb coupling between electrons and the macroscopic electric field generated by LO phonon-induced atomic displacements. Although both transverse optical and LO phonons couple to electrons via non-polar deformation potentials, only LO phonons are considered here due to their dominant impact in polar crystals at elevated temperatures. Contributions from impurity and acoustic phonon scattering are typically negligible in polar semiconductors like perovskites and are hence neglected.

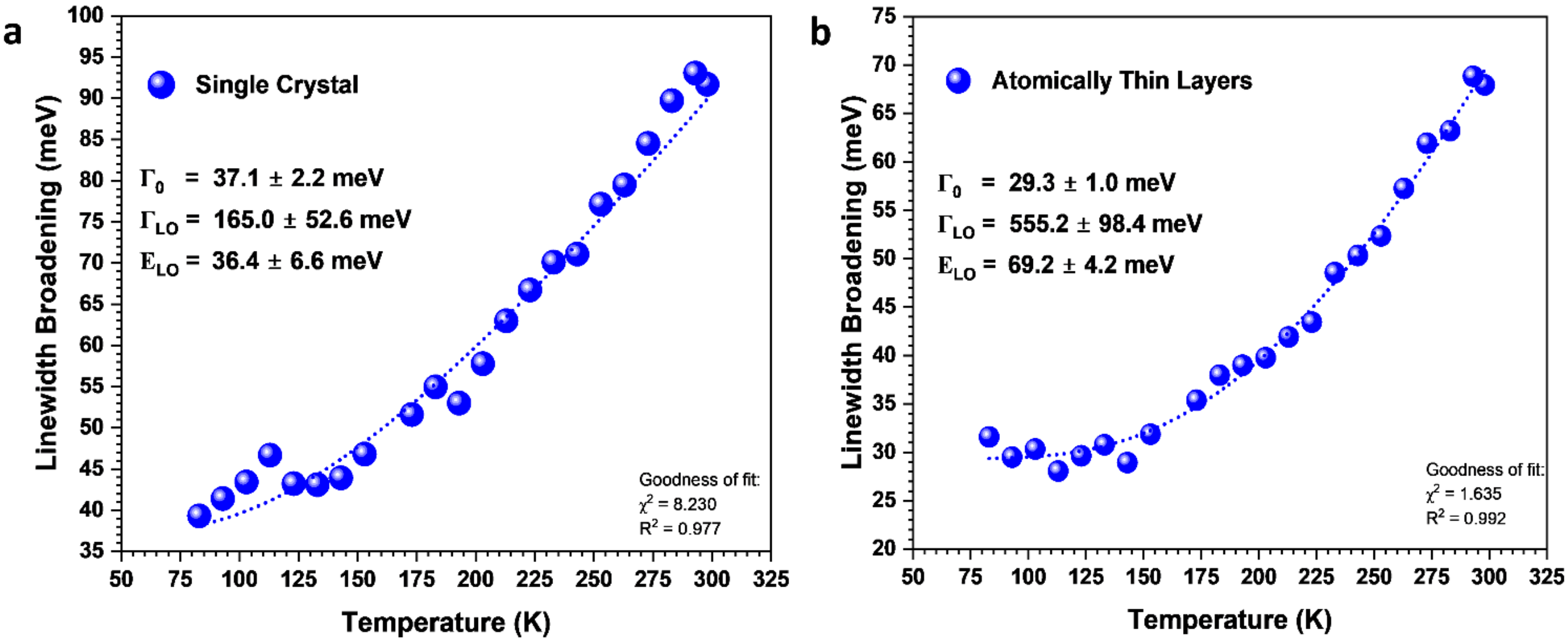

***Figure 14:*** Functional dependence of the PL linewidth $\Gamma$ (FWHM) of $X_1$ emission on temperature using equation (1) to estimate the strength of exciton scattering in **(a)** single crystal **(b)** atomically thin layers.

Our temperature-dependent PL $\Gamma$ (FWHM) linewidth analysis using Bose-Einstein statistical model for single crystals (Supplementary Figure 14a), exfoliated single crystal (Main Figure 4c), and atomically thin layers (Supplementary Figure 14b) yields fits converging to $\Gamma_{ac} \rightarrow 0$. The extracted LO phonon coupling constants $\Gamma_{LO}$ are 165$\pm$52.6 meV, 165$\pm$28.0 meV, and 555.2$\pm$98.4 meV, respectively, indicating notably strong exciton–phonon coupling, with a pronounced enhancement in atomically thin layers of $(PA)_2FAPb_2I_7$, compared to other 2D perovskites[37,38,39]. As observed, the LO-coupling for single crystal and exfoliated single crystal is almost similar. One of the factors affecting this outcome might be that single crystal measurements are not adequate enough to explore the anisotropic-dependent PL properties, considering that it is being averaged across a range of crystal orientations. Hence, exfoliated single crystals with uniform thickness and well-defined edges are more precise and robust to explore exciton-phonon coupling.

## Supplementary Note 12

Various explanations have been proposed to account for the broad and/or multiple emission features in low-dimensional perovskites, which typically include - defect/trap states, self-trapped excitons (STEs), and triplet-state emission. In light of these potential scenarios, we critically analyze our results to determine whether these scenarios are in agreement with our observed data for secondary PL emissions in exfoliated $(PA)_2FAPb_2I_7$ single-crystal sample.

1) The temperature-dependent steady-state PL spectrum indicates a decrease in FWHM of $X_1$ and $X_2$ with a decrease in temperature from 373 K to 83 K (Supplementary Figure 15a). Similarly, the FWHM of $X_5$ decreases (from 438 to 140 meV) as temperature is lowered from 373 K to 203 K. The FWHM slightly increases to 198 meV as the temperature further decreases to 143 K. Below 143 K, the FWHM of $X_5$ remains fairly constant. The FWHM trend with respect to temperature variation does not imply that $X_5$ is a defect/trap-mediated emission, since the characteristic behavior of an emission from a localized trap state typically involves increment in FWHM with decreasing temperature[40]. In conjunction with this, when the full width at half maximum (FWHM) of $X_5$ is fitted as a function of temperature using the Configuration Coordinate model[38], the resulting equation fails to provide a satisfactory fit (Supplementary Figure 15b). Since the Configuration Co-ordinate model proposed by Stadler et al., was fundamentally derived to study the strength for electron–phonon coupling related to emission from defects acting as color centers, the model fails for low-energy $X_5$ emission, thus eliminating the likelihood of $X_5$ originating from defect-mediated emission.

2) The STE trapping/detrapping states can be determined from the temperature dependence of the intensity ratio between the secondary emission and band-edge exciton emission[41,42]. The intensity ratio between low-energy $X_5$ emission and the high-energy $X_1$ band-edge emission plotted in Supplementary Figure 15c, denotes no local maximum found in the low temperature regime. Thus, precluding the plausibility of $X_5$ associated with any STE-mediated emissions.

3) Another interesting possibility is that the secondary emission arises as an outcome of delayed fluorescence (triplet excited state $T_1$ to the singlet excited state $S_1$)) or phosphorescence (spin-forbidden triplet excited state $T_1$ to the singlet ground state $S_0$), which has recently been observed in NEA-based and PEA-based 2D perovskites[43,44]. Typically, the triplet-to-singlet transition is suppressed at higher temperature and becomes more pronounced as the temperature reduces. However, the Low-energy $X_5$ exhibits an exceptionally divergent trend with variation in temperature. Besides, the alkylammonium cations may not have sufficient exciton energy to populate the organic cation's singlet and triplet energy levels. Moreover, the pure-iodide 2D perovskites do not engage in triplet energy transfer as per previous studies[43,44]. Thus, inaccessible singlet and triplet energy levels in $(PA)_2FAPb_2I_7$, negates the feasibility of this scenario.

Based on the preceding discussion, no definitive conclusion could be drawn. Therefore, we delve into a more detailed analysis to investigate the underlying cause of secondary emissions, in particular, the low-energy $X_5$. The detailed discussion can also be found in the main paper.

4) As shown in Main Figure 4(d,e), power-dependent PL measurements at room temperature reveal a superlinear intensity ($\kappa$ = 1.19 $\pm$ 0.09) trend for $X_1$ and almost linear intensity ($\kappa$ = 0.90 $\pm$ 0.05) increment for $X_5$, confirming their intrinsic nature[45]. Whereas at 93 K, $X_1$ exhibits superlinear behaviour ($\kappa$ = 1.39 $\pm$ 0.16) and secondary emissions - $X_3$, $X_4$, $X_5$ exhibits linear power dependence (Main Figure 4(d,e)), retaining its intrinsic nature, and thus negating the possibility of any localized state (STEs/defect states)[45]. Although, $X_2$ emission has a sublinear dependence on power, which indicates formation of localized states.

5) The dominance of the $X_5$ PL emission at elevated temperatures provides strong evidence supporting its assignment as an exciton peak (Main Figure 5a). Notably, above 350 K, $X_5$ remains the sole observable PL peak, indicating its exceptional thermal stability compared to $X_1$ spectral features at elevated temperature. This thermal robustness is characteristic of excitons, whose trapping potential generated by the superlattice preserves their integrity against thermal dissociation, unlike self-trapped excitons (STEs) or defect-related emissions which typically quench rapidly with increasing temperature[46].

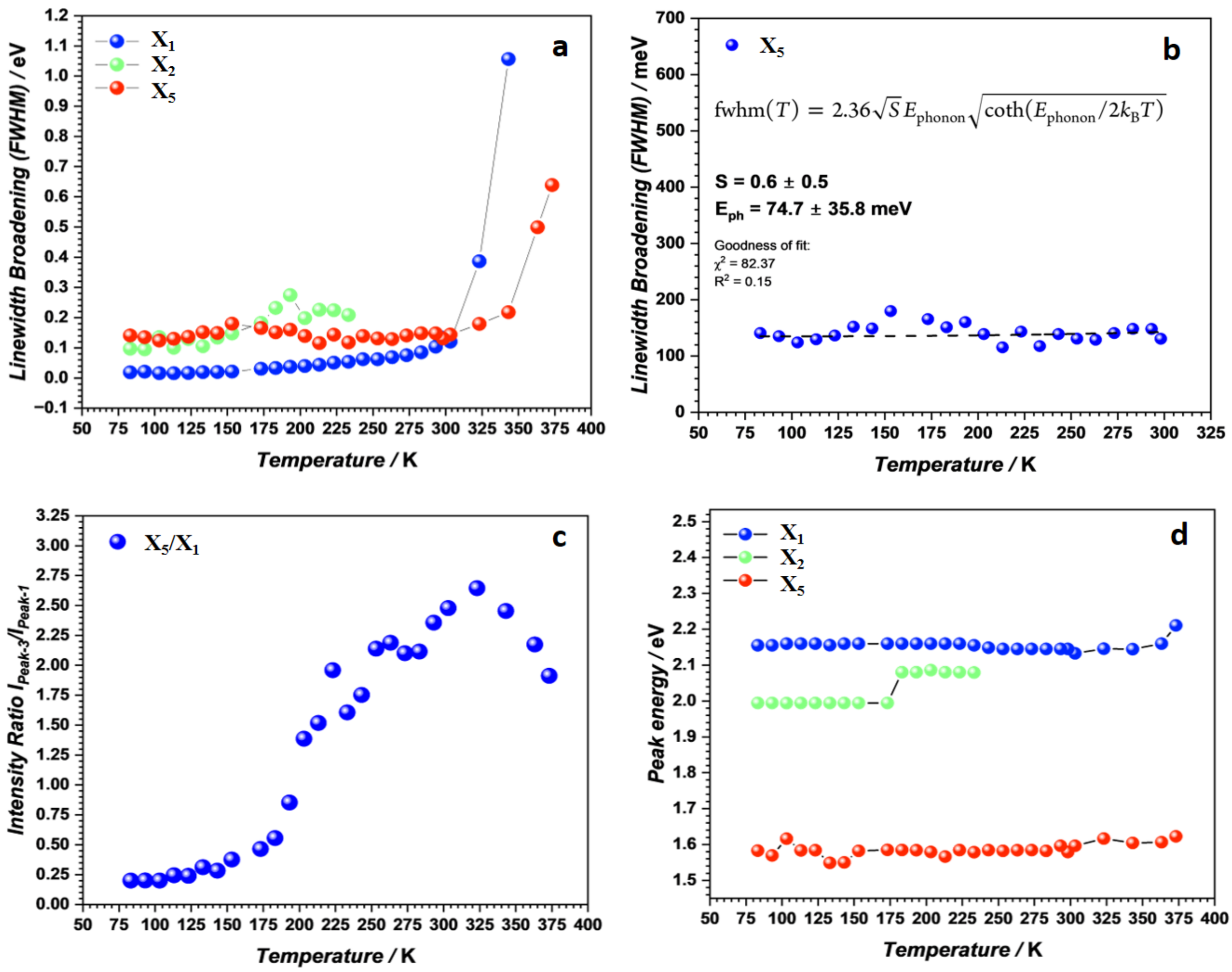

**Figure 15:** **(a)** Variation of Linewidth broadening (FWHM) of $X_1$, $X_2$, $X_5$ as described in main text Figure-4(g) of exfoliated single crystal with temperature. **(b)** Functional dependencies of the PL linewidth $\Gamma$ (FWHM) of the Low-energy $X_5$ emission on temperature using Configuration Co-ordinate model equation, derived to study the strength for electron–phonon coupling related to emission from defects acting as color centers. **(c)** Low and high temperature PL emission. Intensity ratio of the low-energy $X_5$ emission bands with respect to the high-energy $X_1$ band-edge emission. With no local maximum found in the low temperature regime, we cannot associate LEP to STE emission. **(d)** Variation in Energy shift of $X_1$, $X_2$, $X_5$ of exfoliated single crystal with respect to temperature.

## Supplementary Note 13

Polarization-resolved PL measurements further corroborate the excitonic nature of $X_5$. Main Figure 5b shows polarization-dependent anisotropy in peak $X_5$. Defect states or STE emissions do not usually exhibit the strong, stable, and high linear polarization as seen in well-ordered excitonic states like excitons. The linear-polarization can be quantitatively measured using degree of linear polarization (DOLP) parameter (Main Figure 5c) The lower bound of degree of linear polarization (DOLP) for $X_5$ is approximately 0.6, signifying a pronounced polarization-dependent anisotropy. In striking contrast, the lower-energy peak $X_1$ exhibits a negligible DOLP, consistent with isotropic emissions. High DOLP values reflect the inherent anisotropic exciton environment imposed by the potential landscape, governed by the superlattice symmetry and valley-selective optical selection rules. Such a strong polarization anisotropy is generally absent in STEs or defect emissions. While STEs and defect states can exhibit linear polarization in their PL emission, this polarization primarily manifests as variations in the PL intensity with analyzer angle, the peak emission energy remains unchanged with analyzer angle[47]. In contrast, for excitons, the periodic superlattice creates spatially varying potential wells that give rise to multiple excitonic subbands with distinct energies[48]. As the analyzer angle is varied (Main Figure 5d), different exciton states with slightly different energies and polarization selection rules dominate the PL emission. This leads to small but measurable shifts in the PL peak energy with analyzer angle.

In a nutshell, the persistence of $X_5$ emission at room temperature and its significant polarization anisotropy strongly indicate that these secondary PL peaks originate from exciton states localized within the potential generated by moiré superlattice-induced via rotational misalignment twisting of vdW layers. These findings align with recent experimental and theoretical studies demonstrating that moiré excitons form discrete, robust emission features exhibiting unique optical selection rules due to the spatial modulation of the excitonic landscape.

These observations are consistent with the moiré superlattice characteristics inferred from the structural analysis of the crystals, which exhibited rotational misalignment during single crystal growth. The presence of twisted layers gives rise to a moiré superlattice, characterized by a periodic modulation of the local potential landscape with alternating regions of enhanced and reduced potential energy. This moiré superlattice acts as an excitonic confinement framework, whereby moiré excitons get trapped in the potential minima, resulting in emergence of excitonic emissions typically at energies lower than the free excitonic states (Figure 5g-h).

The observation of multiple equidistant PL peaks, may be attributed to the quantum confinement effects induced by the lattice, manifesting as a splitting of the excitonic density of states into discrete quantized minibands[49,50]. These equidistant emissions are indicative of quantized energy levels arising from the spatially periodic potential wells[7]. At lower temperatures, thermal broadening is substantially reduced, enabling the resolution of individual transitions between these quantized minibands, which otherwise merge into broader features at elevated temperatures. At higher and moderate temperatures, increased thermal energy facilitates enhanced exciton mobility, allowing carriers to access and populate potential traps more effectively[51]. This dynamic leads to increased radiative recombination from these localized states, as phonon-assisted hopping over shallow potential barriers becomes more probable. Conversely, at lower temperatures, excitons become strongly localized within the deeper traps, reducing their mobility and favoring non-radiative recombination pathways.

## Supplementary Note 14

The concept of commensurability is pivotal for understanding superlattice formation, especially in complex or non-hexagonal layered materials. In general, commensurability describes the condition in which two periodically stacked lattices achieve an exact match with a certain number of unit cells, resulting in a long-range ordered superlattice. While the commensurability conditions in hexagonal lattices (such as graphene or hBN) are well established, the formalism is readily extended to non-hexagonal systems with arbitrary symmetry and lattice mismatch. For two rotated lattices, the commensurability condition[52] is defined by the existence of integer pairs ($m_1$, $m_2$, $n_1$, $n_2$) satisfying

$$R\left(-\frac{\theta}{2}\right)(m_1 a_1^t + m_2 a_2^t) = R\left(\frac{\theta}{2}\right)\left(n_1 a_1^b + n_2 a_2^b\right) \tag{2}$$

$$where, \qquad R(\varphi) = (cos\varphi \ -sin\varphi \ sin\varphi \ cos\varphi \ )$$

$R\left(\pm\frac{\theta}{2}\right)$ are counter-rotations by half the relative twist angle, and $a_i^{t,b}$ are the primitive lattice vectors of the top (t) and bottom (b) layers, respectively[52]. This equation allows a practical determination of twist angles that yield a common periodic supercell (i.e., commensurate stacking), and those which do not (i.e., incommensurate stacking). In particular, only for discrete, symmetry-allowed twist angles does the atomic unit cell periodically repeat over the entire system, producing a uniform potential modulation. Applying this commensurability condition to our layered $(PA)_2FAPb_2I_7$ perovskite system, we systematically evaluated commensurable and incommensurable twist angles, as shown in Supplementary Table 5. By substituting lattice vectors and iteratively searching for integer solutions, we found that few of the twist angles (θ = 6˚, 10˚, 17˚, 19˚) satisfy the commensurability condition precisely, corresponding to allowed commensurate supercells; while rest twist angles (especially θ = 5-5.5˚) do not, thereby signifying quasi-incommensurate or aperiodic stacking. This result not only validates the non-uniform twist angles observed during crystal growth but also lays the theoretical foundation for the coexistence of distinct superlattice configurations within our samples, depending on subtle variations in rotation and lattice relaxation due to temperature variation.

Our structural investigations, primarily via temperature-dependent diffraction, provide direct evidence supporting this theoretical analysis. The diffraction spot pattern (Main Figure 5e) acquired at low temperature (123 K) reveals sharp, intense, and periodically arranged superlattice reflections. These features are direct signatures of long-range periodic order and structural coherence, as expected for a commensurate superlattice. In contrast, at room temperature (298 K) (Main Figure 5f), the diffraction pattern becomes broadened, diffused, and spatially incoherent, typical of an incommensurate phase in which the lack of strict periodicity results in only short-range order and weak modulation. This structural evolution with temperature signifies a thermally-driven commensurate-incommensurate (C-IC) phase transition[53]. Such C-IC transitions are well-documented in layered oxides, perovskites, and van der Waals systems, where temperature controls the balance between elastic, interlayer, and entropic energies.

This structural transition directly manifests in our temperature-dependent PL measurements. At 123 K, where the diffraction data confirm a commensurate superlattice, the PL spectra exhibit multiple distinct, equidistant emission peaks. These features are interpreted as quantized energy levels described as mini-bands arising from strong potential trapping of excitons, as predicted for uniform, periodic superlattices[7,54]. The emergence of discrete energy levels reflects the spatial periodicity of the potential landscape and robust confinement of quasiparticles, consistent with prior theoretical modeling and other reports on twisted TMD bilayers and 2D perovskites[49,55]. As temperature increases, the system transitions into the incommensurate phase: exciton localization weakens, the PL spectra become broader and less structured, and the mini-band quantization is lost or masked by disorder-induced broadening and increased phonon scattering. This behavior is widely observed in exciton studies across varied materials, where loss of structural commensurability leads to exciton delocalization. Similarly, the sublinear power dependence observed at low temperature (93 K), can be associated with localization of excitons by potentials in a commensurate lattice, and superlinear power dependence at RT, where excitons become more delocalized due to incommensurate structure.

Fundamentally, the C-IC phase transition is governed by the interplay of interlayer commensuration energy (favoring periodicity), lattice elastic energy (imposed by mismatch), and configurational entropy (enhanced at elevated temperatures). At low temperatures, the system minimizes free energy by adopting commensurate stacking, maximizing potential uniformity and structural coherence. At higher temperatures, entropy and dynamic lattice distortions destabilize this periodicity, thus favoring incommensurate or disordered phases, with direct consequences for emergent optical and electronic phenomena. Our findings, therefore, establish a clear linkage between the presence of equidistant, sharp PL peaks and well-ordered diffraction features at low temperature which directly result from commensurate superlattice formation, while their absence or broadening at higher temperature signals the loss of such order due to incommensurability.

**Table 5**: *E*valuated commensurable and incommensurable twist angles from commensurability condition[52] using Pnma structure of $(PA)_2FAPb_2I_7$

| Twist angle | Commensurability | Superlattice structure of T op (m1, m2) and Bottom (n1, n2) layer |
|---|---|---|
| 1° | Incommensurate | - |
| 2° | Incommensurate | - |
| 3° | Incommensurate | - |
| 4° | Incommensurate | - |
| 5° | Incommensurate | - |
| 6° | Commensurate | m1=-1, m2=19, n1=1, n2=19<br>m1=1, m2=-19, n1=-1, n2=-19 |
| 7° | Incommensurate | - |
| 8° | Incommensurate | - |
| 9° | Incommensurate | - |
| 10° | Commensurate | m1=-13, m2=16, n1=-10, n2=18<br>m1=-10, m2=-18, n1=-13, n2=-16<br>m1=10, m2=18, n1=13, n2=16<br>m1=13, m2=-16, n1=10, n2=-18 |
| 11° | Incommensurate | - |
| 12° | Incommensurate | - |
| 13° | Incommensurate | - |
| 14° | Incommensurate | - |
| 15° | Incommensurate | - |
| 16° | Incommensurate | - |
| 17° | Commensurate | m1=-15, m2=8, n1=-12, n2=12<br>m1=15, m2=-8, n1=12, n2=-12 |
| 18° | Incommensurate | - |
| 19° | Commensurate | m1=-12, m2=-2, n1=-12, n2=2<br>m1=-6, m2=-1, n1=-6, n2=1<br>m1=6, m2=1, n1=6, n2=-1<br>m1=12, m2=2, n1=12, n2=-2 |
| 20° | Incommensurate | - |

## References

1. Oh, Y. S. & others. Experimental demonstration of hybrid improper ferroelectricity and the presence of abundant charged walls in (Ca,Sr)3Ti2O7 crystals. *Nat. Mater.* **14**, 407–413 (2015).

2. Smith, K. A., Nowadnick, E. A., Fan, S. & others. Infrared nano-spectroscopy of ferroelastic domain walls in hybrid improper ferroelectric Ca3Ti2O7. *Nat. Commun.* **10**, 5235 (2019).

3. Du, Q. & others. Stacking effects on electron–phonon coupling in layered hybrid perovskites via microstrain manipulation. *ACS Nano* **14**, 5806–5817 (2020).

4. Sung, S. H. & others. Torsional periodic lattice distortions and diffraction of twisted 2D materials. *Nat. Commun.* **13**, 7826 (2022).

5. Cao, Y. & others. Unconventional superconductivity in magic-angle graphene superlattices. *Nature* **556**, 43–50 (2018).

6. Dean, C. R. & others. Hofstadter’s butterfly and the fractal quantum Hall effect in moiré superlattices. *Nature* **497**, 598–602 (2013).

7. Jin, C. & others. Observation of moiré excitons in WSe2/WS2 heterostructure superlattices. *Nature* **567**, 76–80 (2019).

8. Zhang, L., Zhang, X. & Lu, G. Predictions of moiré excitons in twisted two-dimensional organic–inorganic halide perovskites. *Chem. Sci.* **12**, 6073–6080 (2021).

9. Seol, D., Kim, B. & Kim, Y. Non-piezoelectric effects in piezoresponse force microscopy. *Current Applied Physics* **17**, 661–674 (2017).

10. Xue, F. & others. Multidirection piezoelectricity in mono- and multilayered hexagonal α-In2Se3. *ACS Nano* **12**, 4976–4983 (2018).

11. Garduño-Medina, A., Vázquez-Delgado, M., Diliegros-Godines, C., García-Vázquez, V. & Flores-Ruiz, F. Measurements outside resonance with piezoresponse force microscopy. in *AIP Conference Proceedings* vol. 2416 (AIP Publishing, 2021).

12. Gomez, A., Puig, T. & Obradors, X. Diminish electrostatic in piezoresponse force microscopy through longer or ultra-stiff tips. *Appl. Surf. Sci.* **439**, 577–582 (2018).

13. Killgore, J. P., Robins, L. & Collins, L. Electrostatically-blind quantitative piezoresponse force microscopy free of distributed-force artifacts. *Nanoscale Adv.* **4**, 2036–2045 (2022).

14. Kim, S., Seol, D., Lu, X., Alexe, M. & Kim, Y. Electrostatic-free piezoresponse force microscopy. *Sci. Rep.* **7**, 41657 (2017).

15. Rahman, S. & others. Enhanced piezoresponse in van der Waals 2D CuCrInP2S6 through nanoscale phase segregation. *Nanoscale Horiz.* (2025).

16. Hautzinger, M. P., Mihalyi-Koch, W. & Jin, S. A-site cation chemistry in halide perovskites. *Chemistry of Materials* **36**, 10408–10420 (2024).

17. Rahmany, S. & others. The impact of piezoelectricity in low dimensional metal halide perovskite. *ACS Energy Lett.* **9**, 1527–1536 (2024).

18. Frost, J. M. & others. Atomistic origins of high-performance in hybrid halide perovskite solar cells. *Nano Lett.* **14**, 2584–2590 (2014).

19. Li, C. & others. Formability of ABX3 (X = F, Cl, Br, I) halide perovskites. *Acta Crystallogr. B* **64**, 702–707 (2008).

20. Wang, X. & others. Moiré band structure engineering using a twisted boron nitride substrate. *Nat. Commun.* **16**, 178 (2025).

21. Feng, Y. & others. Thickness-dependent evolution of piezoresponses and a/c domains in [101]-oriented PbTiO3 ferroelectric films. *J. Appl. Phys.* **128**, (2020).

22. Feng, Y. & others. Thickness-dependent evolution of piezoresponses and stripe 90 domains in (101)-oriented ferroelectric PbTiO3 thin films. *ACS Appl. Mater. Interfaces* **10**, 24627–24637 (2018).

23. Kelley, K. P. & others. Thickness and strain dependence of piezoelectric coefficient in BaTiO3 thin films. *Phys. Rev. Mater.* **4**, 24407 (2020).

24. Konabe, S. & Yamamoto, T. Piezoelectric coefficients of bulk 3R transition metal dichalcogenides. *Jpn. J. Appl. Phys.* **56**, 98002 (2017).

25. Dong, L., Lou, J. & Shenoy, V. B. Large in-plane and vertical piezoelectricity in Janus transition metal dichalchogenides. *ACS Nano* **11**, 8242–8248 (2017).

26. Wang, Y. & others. Piezoelectric responses of mechanically exfoliated two-dimensional SnS2 nanosheets. *ACS Appl. Mater. Interfaces* **12**, 51662–51668 (2020).

27. Yang, J. & others. Coexistence of piezoelectricity and magnetism in two-dimensional vanadium dichalcogenides. *Physical Chemistry Chemical Physics* **21**, 132–136 (2019).

28. Jiang, X. & others. Strong piezoelectricity and improved rectifier properties in mono- and multilayered CuInP2S6. *Adv. Funct. Mater.* **33**, 2213561 (2023).

29. Guy, I. L., Muensit, S. & Goldys, E. M. Extensional piezoelectric coefficients of gallium nitride and aluminum nitride. *Appl. Phys. Lett.* **75**, 4133–4135 (1999).

30. Calahorra, Y. & Kar-Narayan, S. Piezoelectricity in non-nitride III–V nanowires: Challenges and opportunities. *J. Mater. Res.* **33**, 611–624 (2018).

31. Pham, T. H., Ullah, H., Shafique, A., Kim, H. J. & Shin, Y.-H. Enhanced out-of-plane electromechanical response of Janus ZrSeO. *Physical Chemistry Chemical Physics* **23**, 16289–16295 (2021).

32. Sekhar Muddam, R., Sinclair, J. & Krishnan Jagadamma, L. Piezoelectric charge coefficient of halide perovskites. *Materials* **17**, (2024).

33. Yamashita, Y. & others. A review of lead perovskite piezoelectric single crystals and their medical transducers application. *IEEE Trans. Ultrason. Ferroelectr. Freq. Control* **69**, 3048–3056 (2022).

34. Shi, E. & others. Extrinsic and dynamic edge states of two-dimensional lead halide perovskites. *ACS Nano* **13**, 1635–1644 (2019).

35. Hong, J., Prendergast, D. & Tan, L. Z. Layer edge states stabilized by internal electric fields in two-dimensional hybrid perovskites. *Nano Lett.* **21**, 182–188 (2020).

36. Wright, A. D. & others. Electron-phonon coupling in hybrid lead halide perovskites. *Nat. Commun.* **7**, 11755 (2016).

37. Paritmongkol, W., Powers, E. R., Dahod, N. S. & Tisdale, W. A. Two origins of broadband emission in multilayered 2D lead iodide perovskites. *J. Phys. Chem. Lett.* **11**, 8565–8572 (2020).

38. Lin, M.-L. & others. Correlating symmetries of low-frequency vibrations and self-trapped excitons in layered perovskites for light emission with different colors. *Small* **18**, 2106759 (2022).

39. Ni, L. & others. Real-time observation of exciton–phonon coupling dynamics in self-assembled hybrid perovskite quantum wells. *ACS Nano* **11**, 10834–10843 (2017).

40. Biswas, S. & others. Exciton polaron formation and hot-carrier relaxation in rigid Dion–Jacobson-type two-dimensional perovskites. *Nat. Mater.* **23**, 937–943 (2024).

41. Krahne, R., Lin, M.-L. & Tan, P.-H. Interplay of phonon directionality and emission polarization in two-dimensional layered metal halide perovskites. *Acc. Chem. Res.* **57**, 2476–2489 (2024).

42. Castelli, A. & others. Revealing photoluminescence modulation from layered halide perovskite microcrystals upon cyclic compression. *Advanced Materials* **31**, 1805608 (2019).

43. Blancon, J.-C. & others. Extremely efficient internal exciton dissociation through edge states in layered 2D perovskites. *Science (1979).* **355**, 1288–1292 (2017).

44. Kinigstein, E. D. & others. Edge states drive exciton dissociation in Ruddlesden–Popper lead halide perovskite thin films. *ACS Mater. Lett.* **2**, 1360–1367 (2020).

45. Tongay, S. & others. Defects activated photoluminescence in two-dimensional semiconductors: Interplay between bound, charged and free excitons. *Sci. Rep.* **3**, 2657 (2013).

46. Jin, F. & others. Exciton polariton condensation in a perovskite moiré flat band at room temperature. *Sci. Adv.* **11**, eadx2361 (2025).

47. Xiao, S. & others. Polarization-dependent multiphoton-excited self-trapped emission in alloyed 0D Rb7Bi3Cl16 metal halides via Sb3+ doping. *Small Science* 2500261 (2025).

48. Förg, M. & others. Moiré excitons in MoSe2-WSe2 heterobilayers and heterotrilayers. *Nat. Commun.* **12**, 1656 (2021).

49. Seyler, K. L. & others. Signatures of moiré-trapped valley excitons in MoSe2/WSe2 heterobilayers. *Nature* **567**, 66–70 (2019).

50. Tran, K. & others. Evidence for moiré excitons in van der Waals heterostructures. *Nature* **567**, 71–75 (2019).

51. Rossi, A. & others. Anomalous interlayer exciton diffusion in WS2/WSe2 moiré heterostructure. *ACS Nano* **18**, 18202–18210 (2024).

52. He, Z. & Weng, H. Giant nonlinear Hall effect in twisted bilayer WTe2. *NPJ Quantum Mater.* **6**, 101 (2021).

53. Howieson, G. W. & others. Incommensurate–commensurate transition in the geometric ferroelectric LaTaO4. *Adv. Funct. Mater.* **30**, 2004667 (2020).

54. Wu, F., Lovorn, T. & MacDonald, A. Theory of optical absorption by interlayer excitons in transition metal dichalcogenide heterobilayers. *Phys. Rev. B* **97**, 35306 (2018).

55. Rivera, P. & others. Observation of long-lived interlayer excitons in monolayer MoSe2–WSe2 heterostructures. *Nat. Commun.* **6**, 6242 (2015).